\documentclass[12pt, preprint]{aastex}

\usepackage{graphicx}
\usepackage{amssymb}
\usepackage{epstopdf}
\usepackage{layout}
\usepackage{url}
\usepackage{subfigure}
\slugcomment{Accepted for Publication in {\it The Astrophysical Journal}}

\shorttitle{Time Dependent Hadronic Modeling of FSRQs}
\shortauthors{Diltz et al.}

\begin{document}

\title{Time Dependent Hadronic Modeling of Flat Spectrum Radio Quasars.}

\author{C. Diltz\altaffilmark{1}, \& M. B\"ottcher\altaffilmark{2,1}, G. Fossati\altaffilmark{3}}

\altaffiltext{1}{Astrophysical Institute, Department of Physics and Astronomy, \\
Ohio University, Athens, OH 45701, USA}
\altaffiltext{2}{Centre for Space Research, North-West University, Potchefstroom,
2520, South Africa}
\altaffiltext{3}{Department of Physics and Astronomy, Rice University, Houston, TX 77251, USA}

\begin{abstract}
We introduce a new time-dependent lepto-hadronic model for blazar emission that
takes into account the radiation emitted by secondary particles, such as pions and muons, 
from photo hadronic interactions. Starting from a baseline parameter set guided by a fit 
to the spectral energy distribution of the blazar 3C 279, we perform a parameter study to 
investigate the effects of perturbations of the input parameters to mimic different 
flaring events to study the resulting lightcurves in the optical, X-ray, high energy
(HE: $E > 100$~MeV) and very-high-energy (VHE: $E > 100$~GeV) $\gamma$-rays as well 
as the neutrino emission associated with charged-pion and muon decay. We find that 
flaring events from an increase in the efficiency of Fermi II acceleration will 
produce a positive correlation between all bandpasses and a marked plateau in the HE
$\gamma$-ray lightcurve. We also predict a distinctive dip in the HE lightcurve for 
perturbations caused by a change in the proton injection spectral index. These plateaus 
/ dips could be a tell tale signature of hadronic models for perturbations that lead to 
more efficient acceleration of high energy protons in parameter regimes where pion and
muon synchrotron emission is non-negligible. 
\end{abstract}


\section{Introduction}

Blazars are a subcategory of active galactic nuclei that can be divided into two general classes, 
namely BL Lac objects and Flat Spectrum Radio Quasars (FSRQs). They are typically characterized 
by their rapid variability, superluminal motion and their extreme luminosities, often dominated by 
their $\gamma$-ray emission. These features are considered to be the result of beamed emission 
from a relativistic jet oriented at a small angle with respect to the line of sight \citep{Urry98}. 
The broadband spectral energy distributions (SEDs) of blazars can be typically characterized by two 
broadband, nonthermal peaks that span from the radio to UV wavelengths and from X-rays to high 
energy $\gamma$-rays. It is generally accepted that the first spectral component from radio 
to UV wavelengths is the result of synchrotron radiation of electrons/positrons in a magnetic 
field. For the origin of the second broadband peak, two different paradigms are often invoked, 
collectively referred to as leptonic and hadronic models \citep{Boettcher12}. In the leptonic 
scenario, the high-energy (X-ray -- $\gamma$-ray) emission is due to inverse Compton scattering 
of low-energy photons off the same electrons/positrons. The low-energy target photon fields can
be the synchrotron photons within the emission region (SSC = synchrotron self Compton), or external 
photons (EC = external Compton), which can include the accretion disk \citep{Dermer92,Dermer93}, 
the broad line region \citep[BLR;][]{Sikora94,Blandford95}, or infra-red emitting, warm dust 
\citep{Blazejowski00}. Leptonic models have been quite successful in explaining many features 
in the SEDs and light curves of blazars. 

Hadronic models have also had success in modeling of blazar emission 
\cite[e.g.,][]{MB92,Mannheim93,Mastichiadis95,Mastichiadis05}. 
The hadronic model suggests that the high-energy emission originates 
from the synchrotron emission from a ultrarelativistic protons. The relativistic protons 
interact with the radiation fields within the emission region, producing high energy pions, 
which then decay to produce muons, electrons, positrons, and neutrinos. The pions and their decay
products emit their own radiation (primarily synchrotron) which adds to the broadband spectral 
components in the SEDs of blazars \citep{Aharonian00,Muecke03}. 

As shown in \cite{Boettcher13}, leptonic and hadronic models are generally successful in
reproducing the SEDs of many $\gamma$-ray blazars. Therefore, one needs additional diagnostics
to distinguish which type of model is most applicable to  blazars. The most obvious difference
is the production of TeV -- PeV neutrinos produced only in hadronic models. Additionally, due
to the vastly different acceleration and cooling time scales expected for electrons/positrons
vs. protons, one also expects substantially different variability patterns predicted by the two
types of models. This latter aspect is being studied in detail in this paper. Note that an
alternative discriminant may be the polarization of the high-energy (X-ray -- $\gamma$-ray)
emission of blazars, as discussed in \cite{Zhang13}.

Determining the underlying shapes of the particle distributions that give rise to the broadband
spectral components, is critical to understanding the physics of particle acceleration and cooling in
AGN jets. Simple power law and broken power law proton distributions can be produced through diffusive
shock acceleration when incorporating radiative losses, and such distributions have been invoked in 
hadronic models to explain the emission and subsequent particle cascades that produce high energy 
$\gamma$-rays in blazars \citep[e.g.,][]{Muecke01,Boettcher13}. Second order Fermi acceleration is 
a viable mechanism for producing log-parabola-shaped, curved spectra. The curvature of the spectra 
can give clues to the parameters governing the Fermi II acceleration mechanisms
\citep[e.g.,][]{1Schlickeiser84,Schlickeiser02}. 

Recently, a time dependent hadronic model that considered a Fokker-Planck equation with the incorporation 
of radiative losses, second order Fermi acceleration and the emission produced from the final decay products
of the photo-hadronic interactions was utilized to explain blazar emission \citep{Weidinger14}.
The production rates of final decay products were derived by analytical parametrizations of the
energy distributions for the neutrino, electron/positron and photon distributions from the interactions 
of relativistic protons with the photon fields, \cite{Kelner08}. However, in order for this approach to
be viable, the synchrotron cooling time scales of the intermediate decay products (pions and muons) must
be significantly longer than their decay time scale (in the co-moving frame of the emission region), which
restricts the combination of maximum proton Lorentz factor, $\gamma_{\rm p, max}$, and magnetic field $B$
to $B \, \gamma_{\rm p, max} \ll 5.6 \times 10^{10}$~G \citep{Boettcher13}. If blazar jets are the sites
of the acceleration of ultra-high-energy cosmic rays ($E_p \gtrsim 10^{19}$~eV), then such models are only
applicable in the range of magnetic fields of $B \lesssim 5$~G, substantially lower than usually found in
hadronic modeling of blazars. For higher magnetic fields or maximum proton energies, the synchrotron 
emission from muons and possibly also charged pions becomes non-negligible. At the time of writing, 
no time-dependent hadronic model has been published which incorporates the radiation emitted by the pions 
and muons produced in photo-hadronic interactions.

In this paper, we describe a new, time dependent hadronic model that considers the radiation emitted by all 
secondary products and incorporates Fermi acceleration and self-consistent radiative losses for all particle 
species (including photo-pion production losses for protons). We use this new code to provide a fit to the 
average SED of the FSRQ 3C 279 to determine a baseline parameter set. We then apply a Gaussian perturbation 
to several input parameters (specifically, the magnetic field, the proton injection luminosity, the Fermi-II
acceleration time scale, and the proton injection spectral index) in order to study the resulting light 
curves in the optical, X-rays, HE and VHE $\gamma$-rays as well as neutrinos in the energy range of sensitivity
of IceCube, and study the characteristic variability patterns and cross-correlations between the lightcurves
in the various bandpasses. We describe the model setup in \S \ref{theory}; we present a model fit to the SED 
of the blazar 3C 279, using an equilibrium solution of our code, in \S \ref{spectrum}; we then study the 
light curves resulting from the Gaussian perturbation of the selected parameters in \S \ref{lightcurve}, 
and compute the cross-correlation functions between the various light curves, in \S \ref{correlation}; 
we present a summary and brief discussion in \S \ref{conclusions}.

\section{\label{theory}Model Geometry and the Time Evolution of the Particle Spectra:}

\subsection{\label{theory_general}General Considerations:}

In this section, we describe the assumptions made in our model, the features of the Fokker-Planck 
equations used for the evolution of each particle distribution, and the radiative components 
that they produce. We assume a homogenous, one zone model, where a population of ultra-relativistic 
protons is continuously injected into a spherical region of size $R$, moving along the jet
with a bulk Lorentz factor $\Gamma$, embedded in a homogeneous, randomly oriented magnetic 
field of strength B. The size of the emission region is set by the observed minimum variability 
time scale, $\Delta t_{\rm var}$, through

\begin{equation}
	R = \frac{c \cdot \Delta t_{\rm var} \cdot \delta}{1 + z}
\end{equation}	 

\noindent where $z$ is the redshift to the source and $\delta$ is the Doppler factor. The time 
evolution of the proton distribution is described by a okker-Planck equation that incorporates
cooling due to synchrotron radiation and pion production. The proton distribution interacts 
with the photon fields and generates relativistic pions. The pions 
subsequently decay to produce muons and muon neutrinos. The muons themselves also decay to produce 
electrons, positrons and muon and electron neutrinos. The evolution of each of the particle 
populations is described by their own Fokker-Planck equation.

We assume that the proton injection spectrum takes the form of a power law distribution:

\begin{equation}
	Q_{p}(\gamma) = Q_{p, 0} \gamma^{-q_{p}} H(\gamma; \gamma_{p, min}, \gamma_{p, max})
\end{equation}

\noindent where $H(\gamma; \gamma_{p, min}, \gamma_{p, max})$ denotes the Heaviside function 
defined by $H = 1$ if $\gamma_{p, min} \le \gamma \le \gamma_{p, max}$ and $H = 0$ otherwise. 
The normalization factor for the proton injection spectrum is constrained through the proton 
injection luminosity, $L_{\rm p, inj}$:

\begin{equation}
	Q_{p, 0} = 
		\left\{
			\begin{array}{ll}
				\frac{L_{\rm p, inj}}{V_{b} m_{p} c^{2}} \frac{2-q}{\gamma_{p, max}^{2-q} 
				- \gamma_{p, min}^{2-q}} & \mbox{if $q \ne 2$}, \\
				\frac{L_{\rm p, inj}}{V_{b} m_{p} c^{2} 
				ln(\frac{\gamma_{p, max}}{\gamma_{p, min}})} & \mbox{if $q = 2$}.
			\end{array}
		\right.
\end{equation}					

\noindent where $V_{b}$ denotes the comoving volume of the emission region and $m_{p}$ denotes 
the rest mass of the proton.

All particles can be accelerated or decelerated by gyroresonant interactions with magnetohydrodynamic 
waves. This interaction causes the particle distribution to diffuse in energy, pushing particles to 
higher and lower energies. If the energy density of the plasma waves is small compared to the energy 
density of the magnetic field (quasi-linear approximation), then the diffusion coefficient governing
the momentum diffusion mentioned above, takes the form of a power-law, 

\begin{equation}
	D(\gamma) = K \cdot \gamma^{p}
\end{equation}	

\noindent where the proportionality constant is set by the shock velocity, $v_s$, and the Alfv\'en 
velocity, $v_A$:

\begin{equation}
	K = \frac{1}{(a+2) t_{acc}}	
\end{equation}

\noindent where $a = v_{s}^{2}/v_{A}^{2}$. We invoke a diffusion coefficient with a spectral 
index of $p = 2$ (hard sphere scattering). This allows the acceleration time scale to be 
independent of energy. 

\subsection{\label{theory_pions}Pion Production Templates:}

The protons also interact with the photon fields and produce neutral and charged pions. The total 
proton-photon cross section is divided into separate components, corresponding to different reaction
channels through  which the pions are produced: direct resonances (such as the $\Delta$ resonance), 
higher resonances, direct single-pion production and multi-pion production. We use the prescription 
developed by \cite{Huemmer10} for the photo production rate of pions:

\begin{equation}
	Q_{b}(E_{b}) = \int_{E_{b}}^{\infty} \frac{dE_{p}}{E_{p}} N_{p}(E_{p}) 
	\int_{\frac{\epsilon_{th} m_{p} c^2}{2E_{p}}}^{\infty} d\epsilon n_{\gamma} (\epsilon) R_{b} (x, y)
\end{equation}

\noindent where $N_{p}(E_{p})$ denotes the proton distribution, $n_{\gamma}(\epsilon)$ denotes the 
photon field that the protons interact with as a function of the normalized photon energy $\epsilon
= h \nu / (m_e c^2)$, and $\epsilon_{th} = 294$ (corresponding to an energy of 150~MeV) represents the 
threshold below which the cross sections are zero. The dimensionless variables $x$ and $y$ are given by

\begin{equation}
    x = \frac{E_{b}}{E_{p}}
\end{equation}

\begin{equation}
    y = \frac{E_{p}\epsilon}{m_{p}c^{2}}
\end{equation}

\noindent The {\it response function} $R(x, y)$ in the photo production rate of pions is given by

\begin{equation}
	R_{b} (x, y) = \sum_{IT} R^{IT} (x, y) = \sum_{IT} \frac{1}{2y^{2}} \int_{\epsilon_{th}}^{2y} 
	d\epsilon_{r} \epsilon_{r} \sigma^{IT} (\epsilon_{r}) M_{b}^{IT} (\epsilon_{r}) 
	\delta(x - \chi^{IT} (\epsilon_{r}))
\end{equation}

\noindent and is summed over all interaction channels that make up the proton-photon cross section as a 
function of photon energy in the parent nucleus rest frame, $\sigma^{IT} (\epsilon_{r})$. The functions 
$M_{b}^{IT} (\epsilon_{r})$ and $\chi^{IT} (\epsilon_{r})$ represent the multiplicity of daughter particles 
and the mean energy fraction that is deposited into the daughter particles for a given interaction channel, 
respectively. Evaluating these integrals turns out to be very cumbersome. Therefore, \cite{Huemmer10} suggest
a simplified prescription in which the interactions are split up into separate components that take into 
account the resonances, direct production and multi-production channels and which assumes that the multiplicity 
and deposited mean energy fractions are independent of the interaction energy, $\epsilon_{r}$. The response 
function then simplifies to:

\begin{equation}
	R^{IT}(x, y) = \delta(x - \chi^{IT}) M_{b}^{IT} f^{IT}(y)
\end{equation}

\noindent with

\begin{equation}
	f^{IT}(y) = \frac{1}{2y^{2}} \int_{\epsilon_{th}}^{2y} d\epsilon_{r} \epsilon_{r} 
	\sigma^{IT} (\epsilon_{r})
\end{equation}

With the simplified response function, the photo-production rate of pions can then be written in the more 
compact form:

\begin{equation}
	Q_{b}^{IT} = N_{p}(\frac{E_{b}}{\chi^{IT}}) \cdot \frac{m_{p} c^2}{E_{b}} \int_{\epsilon_{th}/2}^{\infty} 
	dy n_{\gamma}(\frac{m_{p} c^2 \, y \chi^{IT}}{E_{b}}) M_{b}^{IT} f^{IT}(y)
\end{equation}	

\noindent This single integral is easy to evaluate numerically. The photo-production rate of pions now 
depends on the response function, $f^{IT}(y)$, the multiplicities, $M_{b}^{IT}$, and the mean energy 
fraction deposited into the secondary particles, $\chi^{IT}$, for the dominant interaction types. The 
values of the multiplicities and the mean energy deposited for the resonance, direct production and 
multi-pion production as well as the response functions used, are tabulated in \cite{Huemmer10}. 
With the formalism adopted for the pion production rates, the cooling time scale for the proton 
distribution from the production of pions is given by:

\begin{equation}
	t_{cool}^{-1} (E_{p}) = \sum_{IT} M_{p}^{IT} \Gamma^{IT} (E_{p}) K^{IT}
\end{equation}

\noindent where $K^{IT}$ is the inelasticity and $\Gamma^{IT} (E_{p})$ the interaction rate of a 
given interaction type, given by:

\begin{equation}
	\Gamma^{IT} (E_{p}) = \int_{\frac{\epsilon_{th} m_{p} c^2}{2E_{p}}}^{\infty} 
	d \epsilon n_{\gamma} (\epsilon) f^{IT} (\frac{E_{p} \epsilon}{m_{p} c^2})
\end{equation}

\noindent Using this formalism, the rate at which primary protons are lost due to conversion 
into neutrons can be given by the expression:

\begin{equation}
	t_{esc, n}^{-1} (E_{p}) = \sum_{IT, p^{\prime} \ne p} M_{p^{\prime}}^{IT} \Gamma^{IT} (E_{p}) 
\end{equation}
	
\noindent where $p^{\prime}$ denotes the new nucleon created in the photohadronic reaction. 
This formalism also allows the energy-loss term in the proton Fokker-Planck equation due to pion 
production to be defined as:

\begin{equation}
	\dot{\gamma}_{p \gamma} = -E_{p} \cdot t_{cool}^{-1} (E_{p}) = -E_{p} \cdot \sum_{IT} M_{p}^{IT} 
	\Gamma^{IT} (E_{p}) K^{IT}
\end{equation}

\noindent Given that the radiative cooling time scales for protons can be longer than the 
typical dynamical time scale of the expansion of the emission region, we include adiabatic losses 
in our model. Assuming a conical jet with an opening angle of $\theta \sim 1/\Gamma$, the adiabatic
cooling rate is $\dot{\gamma}_{\rm ad} = -3 \, c \, \theta \, \gamma / R$. The full proton Fokker-Planck 
equation that incorporates radiative losses due to synchrotron, adiabatic processes, pion production, 
neutron production as well as stochastic diffusion through the interaction MHD waves reads \citep{Schlickeiser02}:

\begin{equation}
	\frac{\partial n_{p} (\gamma, t)}{\partial t} = \frac{\partial}{\partial \gamma}[\frac{\gamma^{2}}{(a+2) 
	t_{\rm acc}} \frac{\partial n_{p} (\gamma, t)}{\partial \gamma}] - \frac{\partial}{\partial \gamma} 
	(\dot{\gamma} \cdot n_{p} (\gamma, t)) + Q_{p}(\gamma, t) - 
	\frac{n_{p} (\gamma ,t)}{t_{\rm esc}} - \frac{n_{p} (\gamma ,t)}{t_{\rm esc, n}}
\end{equation} 

\noindent where $t_{\rm esc}$ denotes the dynamical escape time scale for the protons which 
we parameterize as a multiple of the light crossing time: $t_{\rm esc} = \eta R/c$ where $\eta \ge 1$. 
The value of $\eta$ is kept as a free parameter. The term $\dot{\gamma}$ denotes the combined 
loss rates on the proton distribution due to adiabatic, synchrotron and pion production processes. 
The synchrotron loss rate is given by

\begin{equation}
	\dot{\gamma}_{\rm rad} = -\frac{4}{3} c \sigma_{T} \frac{u_{B}}{m_{e} c^{2}} (\frac{m_{e}}{m_{p}})^{3} 
	\gamma^{2}
\end{equation}

\noindent where $u_{B}$ is the energy density of the magnetic field. 

\subsection{\label{theory_muons}Pion and Muon Evolution}

The decay time scale of neutral pions (in the pion rest frame) is only $t'_{\rm decay} = 2.8 \times
10^{-17}$~s, and they are not subject to synchrotron losses. Therefore, in our code, neutral pions 
are assumed to decay instantaneously and so, no Fokker-Planck equation needs to be solved. The charged 
pions, however, have a significantly longer half life ($t'_{\rm decay} = 2.6 \times 10^{-8}$~s in the 
pion rest frame), so a separate Fokker-Planck equation has to be considered for the charged pions 
produced in proton-photon interactions:

\begin{equation}
	\frac{\partial n_{\pi} (\gamma, t)}{\partial t} = 
	\frac{\partial}{\partial \gamma}[\frac{\gamma^{2}}{(a+2) t_{\rm acc}}  
	\frac{\partial n_{\pi} (\gamma, t)}{\partial \gamma}] - 
	\frac{\partial}{\partial \gamma} (\dot{\gamma}_{rad} n_{p} (\gamma, t)) 
	+ Q_{\pi}(\gamma, t) - \frac{n_{\pi} (\gamma ,t)}{t_{\rm esc}} - 
	\frac{n_{\pi} (\gamma ,t)}{\gamma t'_{\rm decay}}
\label{FPpion}
\end{equation} 

The only major difference comes from the loss term due to the decay of charged pions with a characteristic 
timescale $t_{\rm decay} = \gamma t'_{\rm decay}$. If the Lorentz factors of the pions are large enough, 
the decay timescale could be of the order of or even larger than the pion synchrotron cooling time scale. 

Charged pions decay to produce muons through the following channels:

\begin{eqnarray}
	\pi^{+} \rightarrow \mu^{+} + \nu_{\mu} \\
	\pi^{-} \rightarrow \mu^{-} + \bar{\nu}_{\mu}
\end{eqnarray}	

\noindent The decay term in the pion Fokker-Planck equation (last term in Eq. \ref{FPpion}) 
serves as the injection function for the muon Fokker-Planck equation. The muons then follow 
their own Fokker-Planck equation that incorporates loss terms due to synchrotron processes 
and diffusive acceleration. The muons can then decay through the following channels:

\begin{eqnarray}
	\mu^{+} \rightarrow e^{+} + \nu_{e} + \bar{\nu}_{\mu} \\
	\mu^{-} \rightarrow e^{-} + \bar{\nu}_{e} + \nu_{\mu}	 
\end{eqnarray}

\noindent to produce separate distributions of electrons, positrons and electron and muon 
neutrinos. In total, we have Fokker-Planck equations for the proton, electron/positron, 
muon, pion, and neutrino distributions that are all coupled to each other and represent 
all particle populations within the emission region. Note that the electron/positron Fokker-Planck
equation contains an additional injection term due to $\gamma\gamma$ absorption and pair production,
which allows us to properly follow the evolution of ultra-high-energy gamma-ray induced pair cascades 
\citep[see, e.g.,][]{Boettcher13}. 

\subsection{\label{theory_radiation}Radiative Contributions:}

Once we know the individual particle spectra, we can compute their radiative output (primarily due to 
synchrotron emission) at any given time step. The synchrotron emission coefficient for a distribution 
$n_i (\gamma)$ of charged particles $i$ within a tangled magnetic field is given by:  

\begin{equation}
	j_{i, syn} (\nu, t) = \frac{1}{4 \pi}\int_{0}^{\infty} d\gamma n_{i} (\gamma, t) \cdot 
	P_{i, \nu} (\gamma)
\end{equation}	

\noindent where the term $P_{i, \nu} (\gamma)$ denotes the synchrotron power per unit frequency 
produced by a single charged particle of species $i$, and can be well approximated by \citep{Boettcher12}:

\begin{equation}
	P_{i, \nu}(\gamma) = \frac{32\pi c}{9\Gamma(4/3)} \, r_{e}^{2}(\frac{m_{e}}{m_{i}})^{2} 
	\cdot u_B \, \gamma^2 \, \frac{\nu^{1/3}}{\nu_{c}^{4/3}} e^{-\nu/\nu_{c}}
\end{equation}

\noindent where $u_{B}$ denotes the energy density of the magnetic field, $r_{e}$ the classical 
electron radius and $m_{i}$ the mass of a particle of species $i$. The critical synchrotron 
frequency $\nu_c$, is given by

\begin{equation}
	\nu_{c} = 4.2 \times 10^{6} \cdot B(G) \cdot(m_{e}/m_{i}) \cdot \gamma^{2} \ Hz
\end{equation}			

The synchrotron spectrum represents one component of the combined photon field that also includes 
the synchrotron-self-Compton and external-Compton radiation of the electrons/positrons (Compton
emission from the heavier particle species is strongly suppressed due to their much higher masses) 
as well as the radiation fields produced by the decay of neutral pions. We then solve a separate 
evolution equation for the combined photon field \citep{Diltz14}:

\begin{equation}
	\frac{\partial n_{ph} (\nu, t)}{\partial t} = \frac{4 \pi}{h \nu} \cdot 
	\sum_{k} j_{k, \nu} (t) - n_{ph} (\nu, t) \cdot (\frac{1}{t_{\rm esc, ph}} + \frac{1}{t_{\rm abs}})
\end{equation}
		
\noindent where the sum is over the all radiation mechanisms and all particle species, 
$t_{\rm esc, ph} = 4 R/3c$ denotes the photon escape time scale and $t_{\rm abs}$ denotes the 
absorption time scale due to synchrotron-self-absorption by electrons and $\gamma\gamma$ absorption. 
The absorption time scale can be defined through the opacity as

\begin{equation}
	t_{\rm abs} = \frac{R}{c \cdot (\tau_{SSA} + \tau_{\gamma \gamma})}
\end{equation}

\noindent where $\tau_{\rm SSA}$ and $\tau_{\gamma \gamma}$ denote the synchrotron-self-absorption and 
$\gamma\gamma$ absorption opacities. 

We utilize the head-on approximation to simplify the differential scattering Compton cross section. 
Using the head-on approximation and assuming that the electron distribution and synchrotron photon 
fields are isotropic in the comoving frame of the emission region, the comoving SSC emission coefficient 
is calculated following \cite{Jones68}. The incorporation of the external radiation fields is implemented 
the same way as in our previous paper on a time-dependent leptonic model \citep{Diltz14}. We compute the 
$\gamma\gamma$ absorption opacity using the prescription of \cite{Dermer09}, and the pair production spectrum
as given in \cite{2Boettcher97}. The produced pair spectrum is added to the solution of the electron/positron 
Fokker-Planck equation at every time step. 

With the combined photon field at every time step, the components that represent the broadband spectral 
energy distribution are then found through:

\begin{equation}
	\nu F_{\nu}^{\rm obs} (\nu^{\rm obs}, t^{\rm obs}) = \frac{h \cdot \nu^{2} \cdot n_{ph} (\nu, t) 
	\cdot \delta^{4} \cdot V_{b}}{4 \pi d_{L}^{2} \cdot t_{\rm esc, ph}}
\end{equation}

\noindent with $\nu^{\rm obs} = \delta \nu$ and $\Delta t^{\rm obs} = \Delta t / \delta$. 

\subsection{\label{theory_neutrinos}Neutrino Emission:}

Our code also takes into account the production rates of electron and muon neutrinos generated
in muon and pion decays following photo hadronic interactions. The neutrino production rate depends 
on the number of charged pions that decay within a given time, $D_{\pi} (E_{\pi})$, which is given 
by the decay term in the pion Fokker-Planck equation:

\begin{equation}
    D_{\pi} (E_{\pi}) = \frac{n_{\pi} (\gamma, t)}{t'_{\rm decay} \, \gamma}
\end{equation}

\noindent With this pion decay rate, the neutrino production rate can be calculated as

\begin{equation}
	Q_{\nu} (E_{\nu}) = \int_{E_{\nu} (1 - r_{M})^{-1}}^{\infty} \frac{dE_{\pi}}{E_{\pi}} \cdot 
	\frac{D_{\pi} (E_{\pi})}{1 - r_{M}}
\end{equation}

\noindent where $r_{M} = m_{\mu}^{2}/m_{\pi}^{2}$. 

The rate of muon decays is governed by the muon Fokker-Planck equation. The calculation of the spectrum 
of neutrinos generated by the decay of charged muons is more difficult than in the case of pion decay,
since the system is a three body decay. We follow the procedure derived in \cite{Barr88} to find the
neutrino production rate for the three-body decay of muons:

\begin{equation}
    Q_{\nu} (E_{\nu}) = \int_{E_{\nu}}^{\infty} dE_{\mu} D_{\mu} (E_{\mu}) \frac{dn}{dE_{\nu}}
    \label{Qnumu}
\end{equation}

\noindent Using the dimensionless scalar variable $m = E_{\nu}/E_{\mu}$, we can recast equation \ref{Qnumu} 
into the form:

\begin{equation}
    Q_{\nu} (E_{\nu}) = \int_{0}^{1} dm \frac{D_{\mu} (E_{\nu}/m)}{m} \cdot \frac{dn}{dm}
\end{equation}

\noindent where $dn/dm$ represents the neutrino production rate in the laboratory frame in terms of the 
dimensionless variable $m$. Assuming that the neutrinos are traveling relativistically, we can cast the 
neutrino distribution function into the following form \citep{Barr88}: 

\begin{equation}
	\frac{dn}{dm} = g_{0} (m) + g_{1} (m)
\end{equation}	

\noindent The scalar functions $g_{0} (m)$ and $g_{1} (m)$ are listed in Table \ref{Neutrino_Functions}
and describe the laboratory-frame distributions of the neutrinos in the relativistic limit. 
Once we have computed the neutrino production rates within the emission region, we determine the expected 
fluxes here on Earth and integrate over the IceCube sensitity range in order to determine the expected 
number of detectable neutrinos. 

\begin{table}[ht]
\centering
\begin{tabular}{ccc}
\hline
& $ g_{0} (m) $ & $ g_{1} (m) $ \\
\hline
$\nu_{\mu}:$ & $\frac{5}{3} - 3m^{2} + \frac{4}{3}m^{3}$ & $\frac{1}{3} - 3m^{2} + \frac{8}{3}m^{3}$ \\
$\nu_{e}:$ & $2 - 6m^{2} + 4m^{3}$ & $-2 +12m - 18m^{2} + 8m^{3}$ \\
\hline
\end{tabular} \\ 
\caption{Laboratory-frame electron and muon neutrino distribution functions 
\citep[from][]{Barr88}}
\label{Neutrino_Functions}
\end{table}

\begin{figure}[ht]
\begin{center}
\includegraphics[height=0.35\textheight]{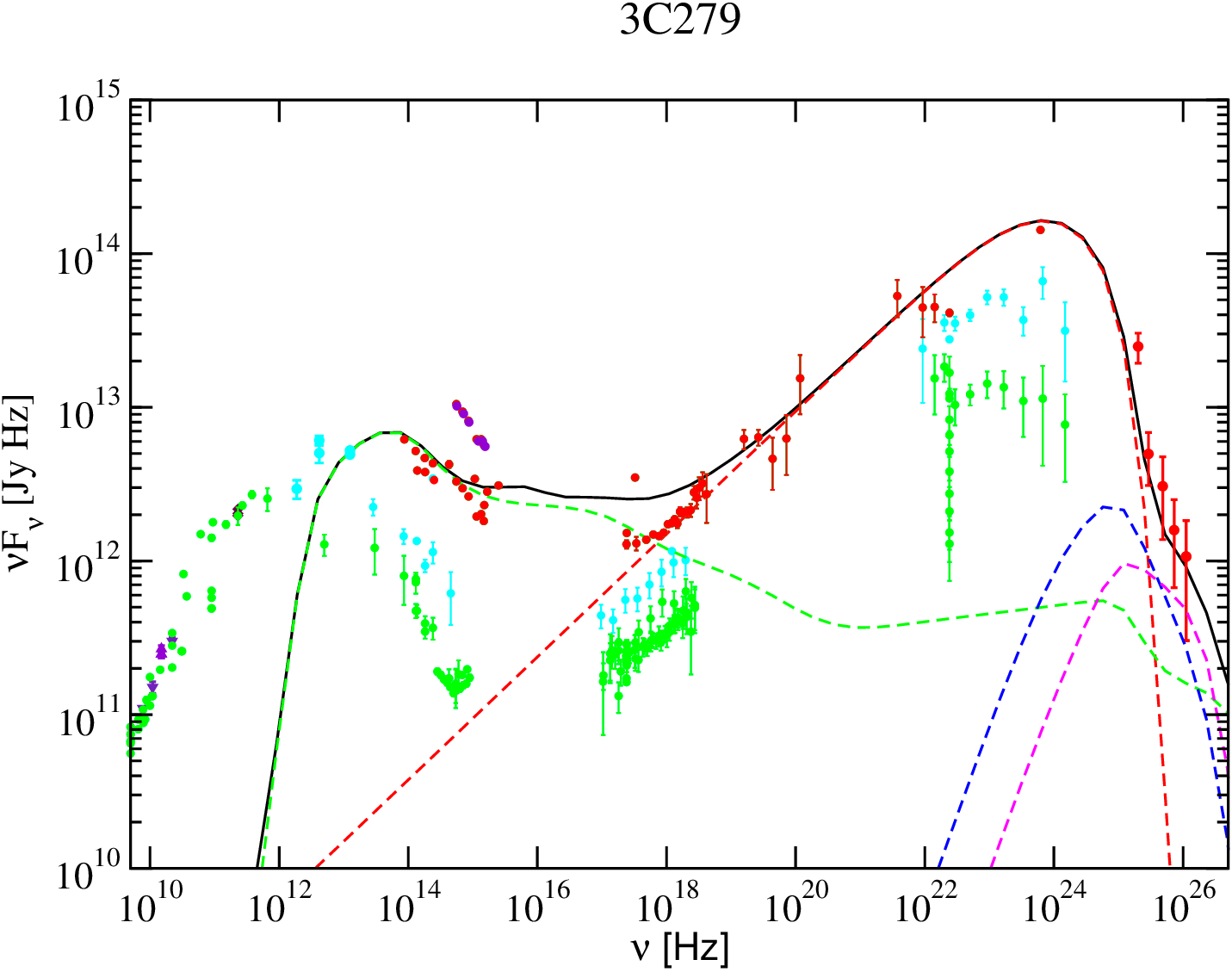}
\caption{Equlibrium fit to the SED of 3C279. The high-state data points included in the fit are plotted
in red; additional, archival data are plotted in other colors \citep[data from][]{Abdo10}. The model curves 
are: green dashed = synchrotron emission from electrons/positron; red dashed = proton synchrotron; blue dashed
= muon synchrotron; magenta dashed = pion synchrotron; black solid = total spectrum. }
\label{3C279fit}
\end{center}
\end{figure}

\section{\label{spectrum}Application to the FSRQ 3C279}

3C 279 $(z = 0.538)$ was the first $\gamma$-ray blazar discovered using the Compton Gamma Ray Observatory and 
has been the target of several multifrequency campaigns \citep[e.g.,][]{Maraschi94,Hartman96,Hartman01,Ballo02}. 
3C 279 has been classified as a flat spectrum radio quasar given the location of its synchrotron peak in the 
infrared. Many observational properties of 3C 279 have been well measured, including the accretion disk luminosity 
\citep{Hartman01}, the bolometric luminosity of the broad line region \citep{Xie08}, the minimum variability 
time scale \citep{Boettcher07} and the apparent superluminal speed of relativistic jet components \citep{Hovatta09}. 
3C 279 is one of only three FSRQs detected in VHE $\gamma$-rays by ground-based Cherenkov Telescope facilities.
Specifically, 3C 279 was detected by the Major Atmospheric Gamma-Ray Imaging Cherenkov (MAGIC) Telescope
during an exceptional $\gamma$-ray flaring state in 2006 \citep{Albert08}. \cite{Boettcher09} have
pointed out that this VHE detection, in combination with the rest of the SED and known variability
properties of 3C 279, presents a severe challenge to single-zone leptonic jet models, and suggest
a hadronic scenario as a viable alternative. For this reason, We choose this well-known blazar as 
a representative of $\gamma$-ray bright blazars in which hadronic processes might be important.

We performed a parameter study to provide a rough fit to the average SED of 3C279 \citep[as presented 
in][]{Abdo10}, by running our time-dependent leptohadronic model code with time-independent input 
parameters and waiting for all particle and photon spectrum solutions to relax to an equilibrium. 
To obtain these equilibrium solutions, we set the size of the time step in our code initially to 
$10^{7}$~s. This time step size is considerably longer than the time scales for all loss terms, 
acceleration terms and escape terms in all particle Fokker-Planck equations. The implicit Crank-Nichelson 
scheme used to numerically solve the Fokker-Planck equations guarantees that the simulation converges
to a stable solution after a few time steps. 

Given the number of input parameters in our model (see Table \ref{parameters}), it is important 
to independently constrain as many parameters as we can from observational data. For 3C279, we 
have the following observational parameters \citep[see][for references to the observational 
data]{Boettcher13}: $z = 0.536$, $\beta_{\perp, \rm app} = 20.1$ (the apparent transverse velocity of 
individual jet components, normalized to the speed of light), $\Delta t_{\rm var} \sim 2$~d, 
$L_{\rm disk} = 2.0 \times 10^{45}$~erg~s$^{-1}$, and $L_{\rm BLR} = 2.0 \times 10^{44}$~erg~s$^{-1}$. 
The superluminal motion speed sets a lower limit to the bulk Lorentz factor, $\Gamma > 21$. The observing 
angle is set by using the relation $\theta_{\rm obs} = 1/\Gamma$ so that $\delta = \Gamma$. 
From the variability time scale, we can constrain the location of the emission region 
along the jet, $R_{\rm axis} \sim 2 \, \Gamma^{2} \, c \, t_{v} / (1 + z) \approx 10^{18} \ cm$. 
With the luminosity of the broad line region, we can determine the characteristic size of the BLR using the 
luminosity-radius relation \citep{Bentz13}. The mass of the supermassive black hole in 3C 279 is 
constrained through the measured bolometric luminosity of the broad line region and is 
found to be $(4-8) \times 10^{8} M_{sol}$ \citep{Woo02}. With the mass of the black hole and 
the accretion disk luminosity, we can then constrain the Eddington ratio for the accretion disk emission. 
We approximate the BLR spectrum as an isotropic (in the AGN rest frame) thermal blackbody with a 
characteristic temperature of $5.0 \times 10^{3} \ K$, which has been shown by \cite{Boettcher13} 
to yield Compton spectra that are virtually indistinguishable from spectra using more complicated 
BLR reprocessing geometries.

Within the parameter constraints listed above, we perform a "fit by eye" to find suitable values 
for the remaining parameters. In the context of most hadronic modeling, the X-ray to soft and 
intermediate $\gamma$-ray emission from FSRQs can be best explained by proton synchrotron radiation. 
Thus, the X-ray through HE $\gamma$-ray spectrum informs our choice of the proton injection luminosity, 
spectral index, and maximum proton energy. The VHE $\gamma$-ray emission detected by MAGIC \citep{Albert08}
appears to constitute a separate radiation component beyond the {\it Fermi}-LAT $\gamma$-ray spectrum,
and we here suggest that this component may be provided by muon- and pion-synchrotron radiation,
which our code is uniquely able to handle in a time-dependent fashion. By chosing $B \gamma_{\rm p, max} 
\gtrsim 5 \times 10^{10} \ G$, our simulations will be in a parameter regime in which muon and pion
synchrotron is expected to make a significant contribution to the $\gamma$-ray emission. A full 
list of parameters which yield a satisfactory representation of the SED of 3C 279, is given in 
Table \ref{parameters}.

\begin{table}[ht]
\centering
\begin{tabular}{ccc}
\hline
Parameter & Symbol & Value \\
\hline
$ Magnetic \ field $ & $ B $ & $150$~G \\ 
$ Radius \ of \ emission \ region $ & $ R $ & $8.69 \times 10^{15}$~cm \\
$ Constant \ multiple \ for \ escape \ time \ scale $ & $\eta$ & 6.0 \\
$ Bulk \ Lorentz \ factor $ & $\Gamma$ & 21 \\
$ Observing \ angle $ & $\theta_{\rm obs}$ & $4.76 \times 10^{-2}$~rad \\
$ Minimum \ proton \ Lorentz \ factor $ & $ \gamma_{\rm p, min} $ & $ 1.0 $ \\
$ Maximum \ proton \ Lorentz \ factor $ & $ \gamma_{\rm p, max} $ & $ 4.5 \times 10^{8} $ \\
$ Proton \ injection \ spectral \ index $ & $q_{p}$ & 2.2 \\
$ Proton \ injection \ luminosity $ & $L_{\rm p, inj}$ & $3.5 \times 10^{46}$~erg~s$^{-1}$ \\
$ Minimum \ electron \ Lorentz \ factor $ & $ \gamma_{\rm e, min} $ & $ 5.1 \times 10^{2} $ \\
$ Maximum \ electron \ Lorentz \ factor $ & $ \gamma_{\rm e, max} $ & $ 1.0 \times 10^{4} $ \\
$ Electron \ injection \ spectral \ index $ & $q_{e}$ & 3.2 \\
$ Electron \ injection \ luminosity \ $ & $L_{\rm e, inj}$ & $7.8 \times 10^{41}$~erg~s$^{-1}$ \\
$ Supermassive \ black hole \ mass $ & $M_{\rm BH}$ & $6.0 \times 10^{8} \, M_{\odot}$ \\
$ Eddington \ ratio $ & $l_{\rm Edd}$ & $1.18 \times 10^{-2}$ \\
$ Blob \ location \ along \ the \ jet \ axis $ & $R_{\rm axis}$ & 0.279~pc \\
$ Radius \ of \ the \ BLR $ & $R'_{\rm ext}$ & 0.071~pc \\
$ Energy \ density \ of \ the \ BLR \ in \ comoving \ frame $ & $u'_{\rm ext}$ & $3.68 \times 10^{-4}$~erg~cm$^{-3}$ \\
$ Blackbody \ temperature \ of \ BLR $ & $T_{\rm BB}$ & 5000~K \\
$ Ratio \ between \ the \ acceleration \ and \ escape \ time \ scales $ & $ t_{\rm acc}/t_{\rm esc} $ & $32.5$ \\
\hline
\end{tabular}
\caption[]{\label{parameters}Parameter values used for the equilibrium fit to the SED of
3C279.}
\end{table}

With this set of baseline parameters, the broadband SED of 3C 279 can be reproduced quite well, 
as shown in Figure \ref{3C279fit}. The infrared to UV portion of the SED is fitted by synchrotron 
radiation from primary electrons. The X-ray to GeV $\gamma$-ray emission in our model SED is 
dominated by proton synchrotron radiation. The VHE $\gamma$-ray spectrum, as measured by MAGIC, 
is best explained by a combination of synchrotron radiation from the primary protons and secondary 
muons and pions generated via photo-pion production. We note that the proton synchrotron component 
slightly overshoots the Fermi data points. This is reasonable since the {\it Fermi}-LAT spectrum 
represents a long-term averaged high-state, while the MAGIC detection corresponds to an exceptional, 
short-term flaring event during which no GeV $\gamma$-ray observatory was operating, but one may 
expect that the HE $\gamma$-ray flux at that time was larger than the {\it Fermi}-LAT high-state 
flux presented in \cite{Abdo10} and shown in Figure \ref{3C279fit}. The radio emission from our 
model simulation is synchrotron-self-absorbed and therefore underpredicts the observed radio 
flux from 3C279. This suggests that the observed radio emission likely originates in more 
extended regions of the jet, beyond the radiative zone considered in our model.

\begin{figure}[ht]
\begin{center}
\includegraphics[height=0.35\textheight]{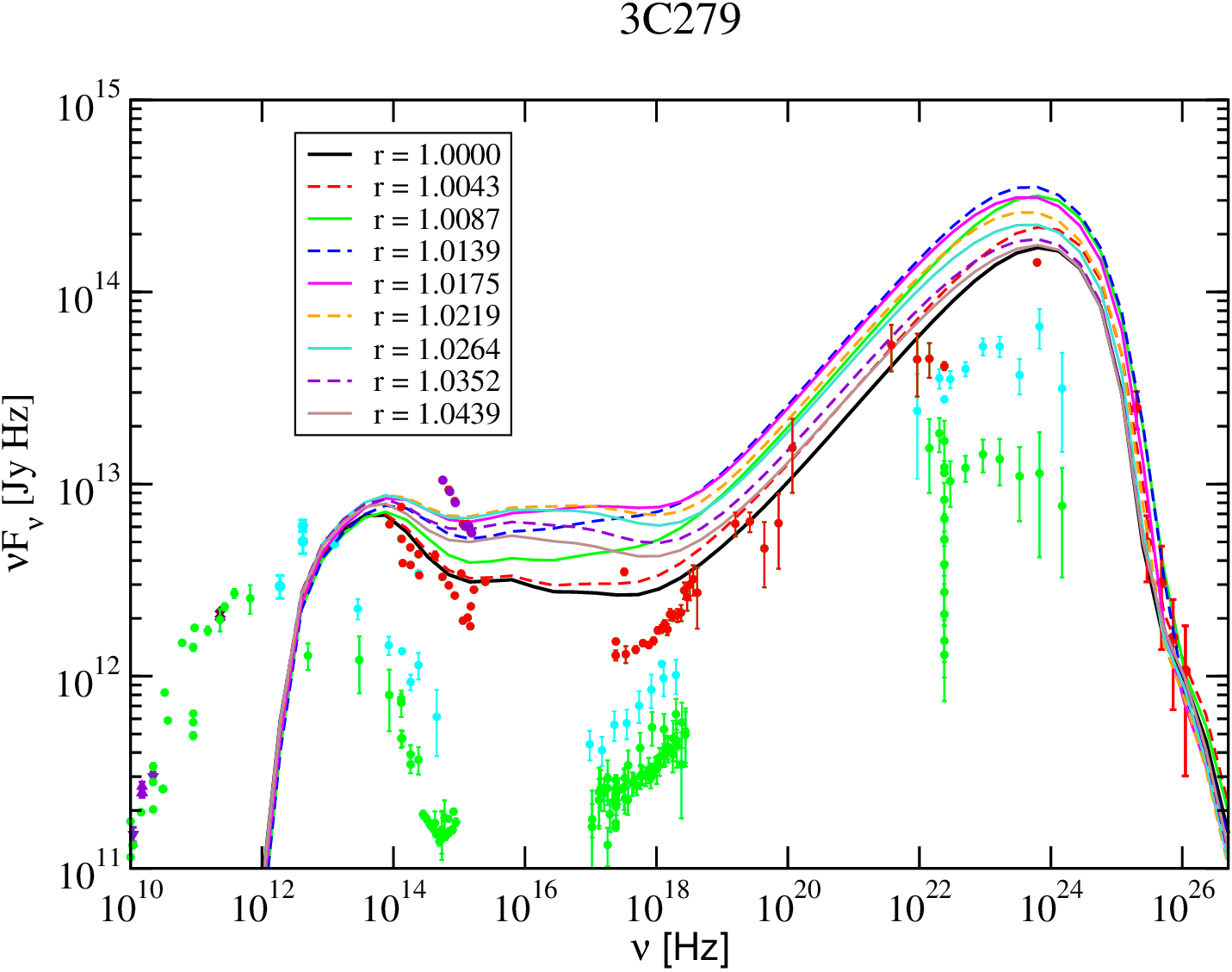}
\caption{Time evolution of the model SEDs for the case of the magnetic field perturbation. The times
are parameterized through $r = (t_{e} + \Delta t)/t_{e}$, where $t_{e}$ denotes the time when the 
perturbation is switched on in the observer's frame.}
\label{BSEDs}
\end{center}
\end{figure}

In our simulation, the jet is --- to within a factor of a few --- in approximate equipartition 
between the powers carried in magnetic fields and in kinetic energy of particles: The power carried 
along the jet in the form of magnetic field (i.e., the Poynting Flux) is determined by

\begin{equation}
    L_{B} = \pi R^{2} \Gamma^{2} \beta_{\Gamma} c \frac{B^{2}}{8 \pi}
\end{equation}

\noindent which, for our baseline fit to 3C279, yields $L_{B} = 2.8 \times 10^{48}$~erg~s$^{-1}$. 
The particle kinetic luminosities in the observer's frame are calculated from the equilibrium
particle distributions $n_i (\gamma_i)$ as

\begin{equation}
    L_{i} = \pi R^{2} \Gamma^{2} \beta_{\Gamma} c \, m_{i} c^{2} \int\limits_{1}^{\infty}
    d\gamma_i \, n_i (\gamma_i) \, \gamma_i 
\end{equation}

\noindent where $i$ denotes the particle species considered. From numerically integrating the solution 
to the Fokker-Planck equation for both the proton and electron/positron distributions when equilibrium 
is reached, we find that the corresponding particle kinetic luminosities are $L_{p} = 9.7 \times 
10^{48}$~erg~s$^{-1}$ and $L_{e} = 3.5 \times 10^{43}$~erg~s$^{-1}$. With these values, the partition 
parameter between the combined particle kinetic luminosity and the power carried by the magnetic 
field, $\epsilon_B \equiv L_{B}/L_{\rm kin}$, where $L_{\rm kin} = L_{e} + L_{p}$, is then 
$\epsilon_B \approx 0.29$. Our value for $L_p$ is similar to the values usually required by most
previously published hadronic model interpretations of FSRQ SEDs. However, previously published 
works usually require parameters far out of equipartition. For example, in \cite{Boettcher13}, 
$L_B / L_p = 7.9 \times 10^{-3}$ for their fit to the SED of 3C279, while our model produces a 
reasonable representation of the same SED with a jet near equipartition. This might be a consequence 
of the higher radiative efficiency in the parameter regime chosen here, with the inclusion of 
secondary muon and pion synchrotron radiation.

\begin{figure}[ht]
\begin{center}
\includegraphics[height=0.35\textheight]{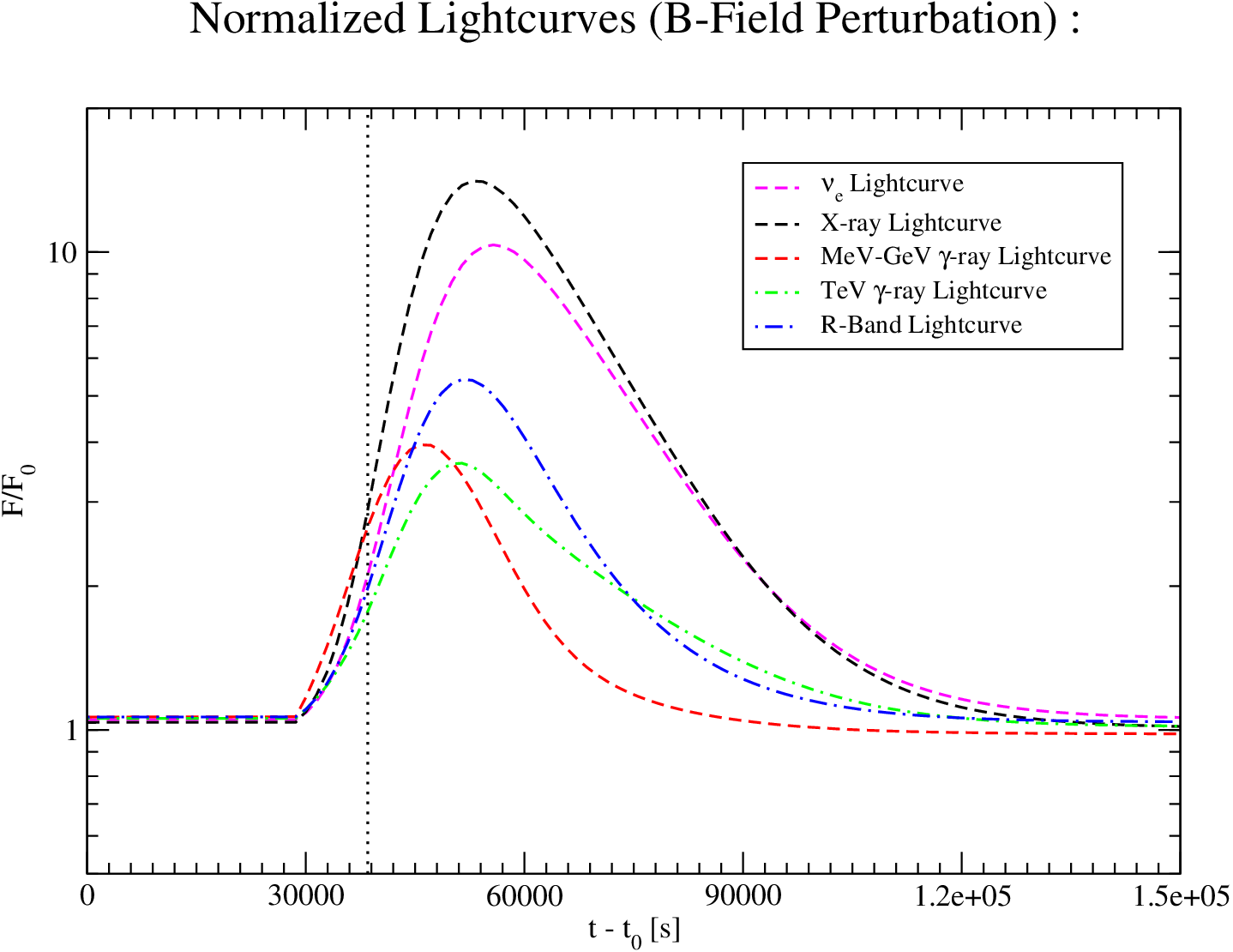}
\caption{Normalized light curves in optical, X-rays, HE and VHE $\gamma$-rays and neutrino flux,
for the magnetic field perturbation as illustrated in Fig. \ref{BSEDs}.}
\label{Blightcurve}
\end{center}
\end{figure}

\begin{figure}[ht]
\begin{center}
\includegraphics[height=0.35\textheight]{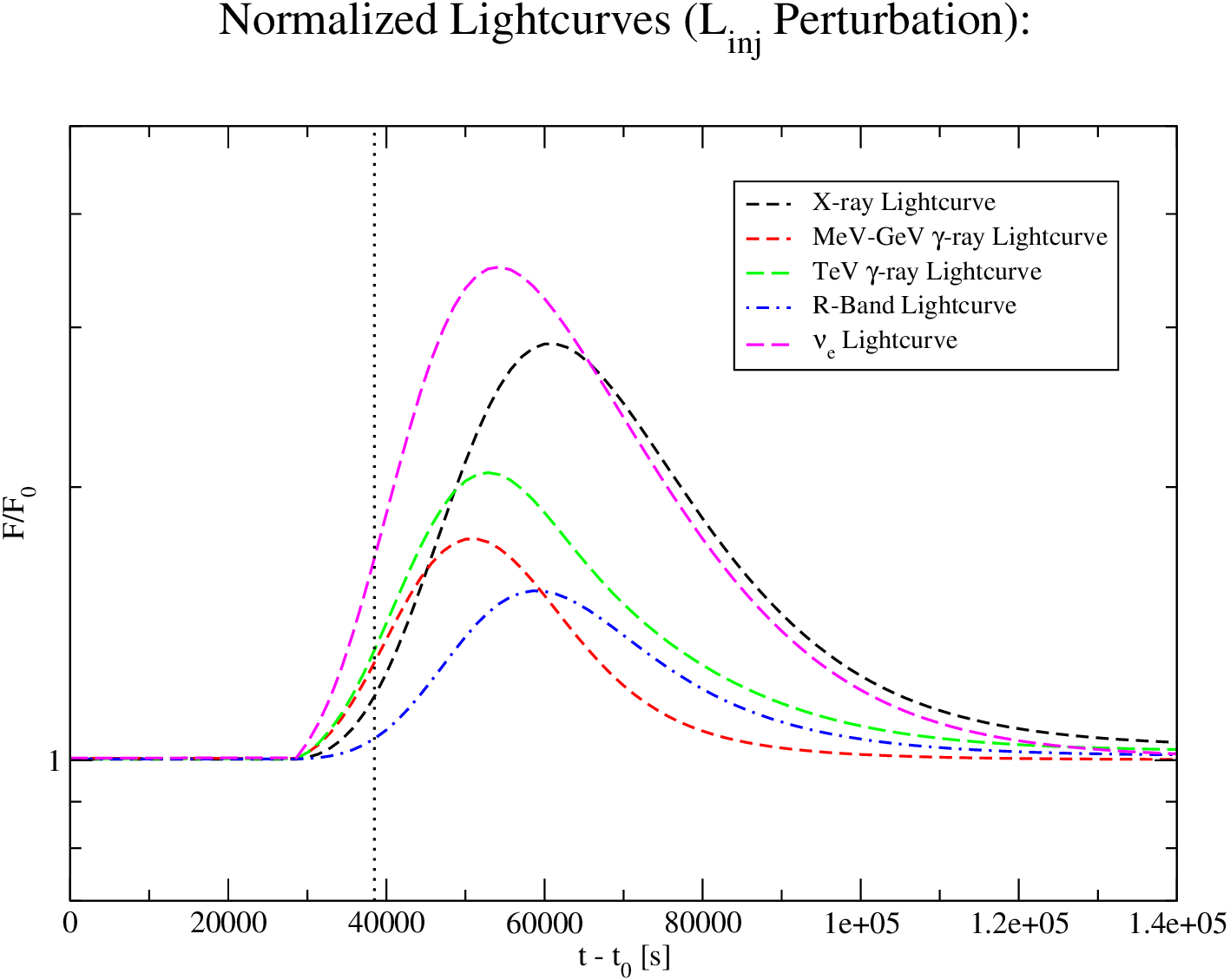}
\caption{Normalized light curves for the proton injection luminosity perturbation.}
\end{center}
\end{figure}

\section{\label{lightcurve}Simulated Lightcurves}

We use the parameter set from our steady-state fit to the SED of 3C 279 as a baseline model from which
we start out to apply perturbations to a few parameters in order to investigate the effects 
of these perturbations on the resulting multiwavelength light curves. Once the model has reached equilibrium
as described in the previous section, we modify the time step to $\Delta t = 2.0 \times 10^{4}$~s. 
This allows us to resolve light curve patterns on time scales characteristic for cooling effects of the 
relativistic protons. However, we are unable to diagnose the shorter-term variability, potentially caused
by the radiative cooling of high energy electron-positron pairs generated from the decay of charged mesons, 
since their cooling time scales are significantly shorter than the size of the time step selected for these 
simulations. Note, again, that the implicit Crank-Nicholson scheme implemented for the solution of the 
Fokker-Planck equations guarantees that a stable solution for the electrons/positrons, muons, and pions
is obtained even if the time step is longer than the radiative coolimg time scale. Decreasing the size 
of the time step would allow us to probe variability on those shorter time scales, but extending such
simulations to time scales of the order of the proton cooling time scales would require prohibitively
long simulation times. 

\begin{figure}[ht]
\begin{center}
\includegraphics[height=0.35\textheight]{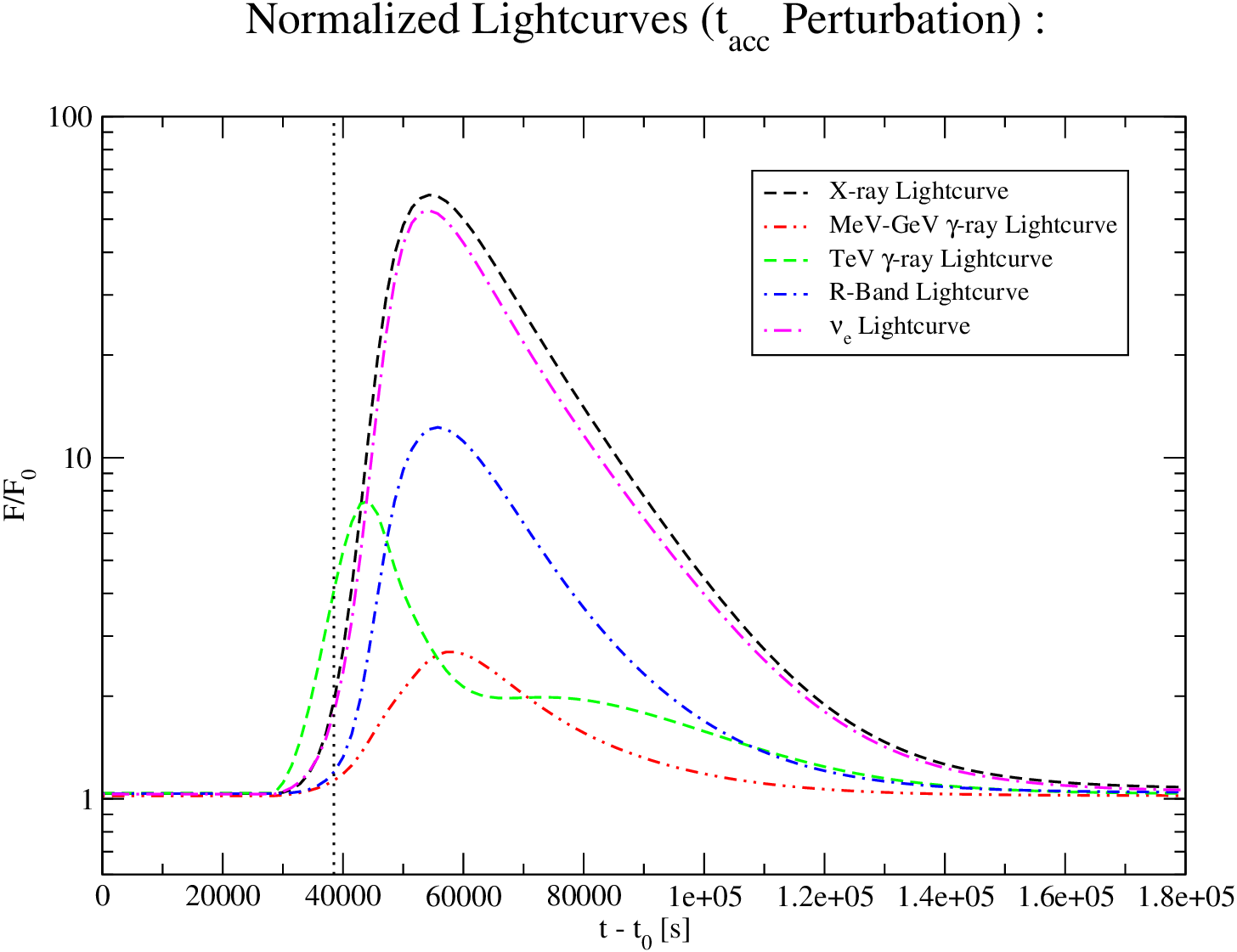}
\caption{Normalized light curves for the acceleration efficiency perturbation.}
\end{center}
\end{figure}

After the simulation has reached equilibrium, one of the input parameters ($B$, $L_{\rm p, inj}$, 
$t_{\rm acc}$, $q_{\rm p}$) is modified in the form of a Gaussian perturbation in time. From the 
outputs produced in the simulations, we integrate the light curves in the following bandpasses: 
optical (R-band), X-ray (0.1~keV -- 10~keV), HE $\gamma$-rays (20~MeV -- 300~GeV) and VHE $\gamma$-rays 
(30~GeV -- 100~ TeV). The magnetic field perturbation is given by

\begin{equation}
B(t) = B_0 + K_{B} \cdot e^{-(t - t_{0})^{2}/2\sigma^{2}}
\label{Bmodification}
\end{equation}

\noindent where $B_{0} = 150$~G denotes the equilibrium value for the magnetic field, 
$K_{B} = 250$~G parametrizes the amplitude of the perturbation, and $t_{0}$ and $\sigma$ 
specify the time when the perturbation reaches its peak and the characteristic time scale 
of the perturbation, respectively. The chosen perturbation for the proton injection luminosity 
has the same functional form,

\begin{equation}
L_{\rm inj} (t) = L_{\rm inj, 0} + K_{L} \cdot e^{-(t - t_{0})^{2}/2\sigma^{2}}
\label{Lmodification}
\end{equation}
	 
\noindent where $L_{\rm inj, 0} = 3.5 \times 10^{46}$~erg~s$^{-1}$ is the equilibrium proton 
injection luminosity and $K_{L} = 0.3 \times L_{\rm inj, 0}$ is the amplitude of the 
perturbation. The perturbation of the acceleration time scale is chosen in such a way that 
the acceleration time scale decreases to a minimum during the peak of the perturbation. This 
is achieved with the following parametrization:

\begin{equation}
t_{\rm acc} (t) = \frac{t_{\rm acc, 0}}{1 + K_{t} \cdot e^{-(t - t_{0})^{2}/2\sigma^{2}}}
\label{taccmodification}
\end{equation}

\noindent where $t_{\rm acc, 0}$ is the equilibrium value of the acceleration time scale and $K_{t} = 14$ 
characterizes the amplitude of the perturbation. We also include a perturbation of the proton spectral 
index such that a flare is caused by a temporarily harder proton spectral index:

\begin{equation}
	q_{p} (t) = q_{p, 0} - K_{q} \cdot e^{-(t - t_{0})^{2}/2\sigma^{2}}
\end{equation}

\noindent where $q_{p, 0}$ denotes the equilibrium value for the proton spectral index and 
$K_{q} = 1.0$ denotes the strength of the perturbation. For all four perturbations, we choose a width of 
$\sigma = 1.0 \times 10^5$~s, corresponding to approximately 10 light-crossing time scales through the
emission region, in the co-moving frame. 

\begin{figure}[ht]
\begin{center}
\includegraphics[height=0.35\textheight]{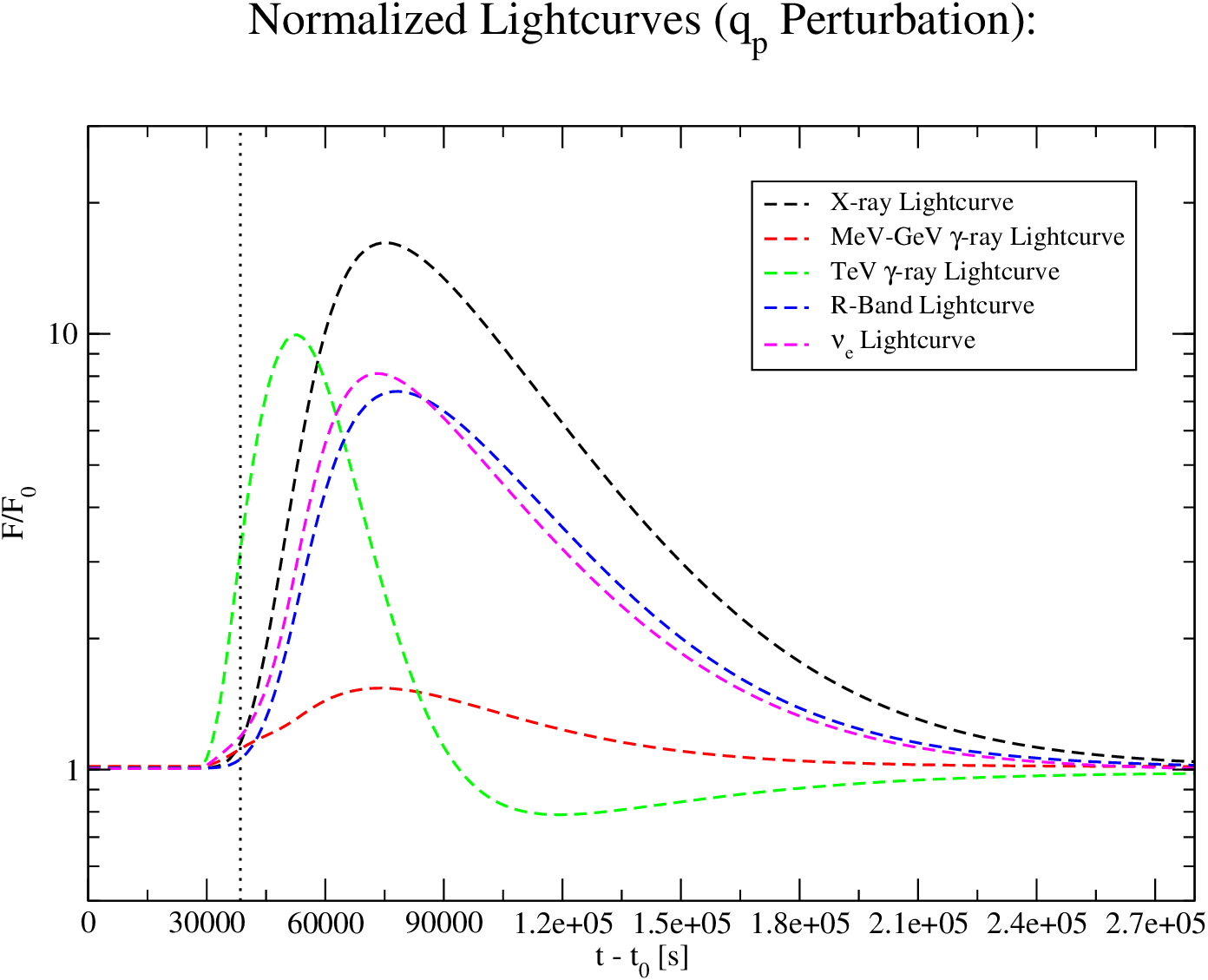}
\caption{Normalized light curves for the proton spectral index perturbation.}
\label{qlightcurve}
\end{center}
\end{figure}

For the example of the modification of the magnetic field, Figure \ref{BSEDs} illustrates snap-shot
SEDs at various times throughout the simulation. The corresponding light curves (normalized to the 
respective peak fluxes) extracted from our time-dependent simulations are shown in figures \ref{Blightcurve} 
to \ref{qlightcurve}.

A temporary increase in the magnetic field obviously leads to a marked increase in the proton synchrotron 
(primarily HE $\gamma$-rays) and electron-synchrotron (IR -- optical) spectral components. The corresponding 
increase of the synchrotron-photon energy density leads to a larger pion-production (and subsequent 
pion- and muon-decay) rate. The resulting pions and muons are also subjected to the increased 
magnetic field, thus strongly increasing the contribution of muon and pion synchrotron to the 
SED. This increase in synchrotron emission from secondary particles leads to a distinct VHE $\gamma$-ray 
flare, slightly delayed with respect to the primary proton-synchrotron (HE $\gamma$-ray) flare. After 
the secondary particles have decayed to electrons and positrons, those secondaries also cool via 
synchrotron emission, producing a marked flare in the X-ray bandpass, with a very short (on the 
electron-synchrotron cooling time-scale) delay with respect to the VHE $\gamma$-ray flare. The 
enhanced pion and muon decay rates also lead to an increased neutrino flux, approximately coincident
with the secondary electron/positron synchrotron (X-ray) flare. 

A perturbation in the proton injection luminosity causes the proton synchrotron emission to increase 
producing primarily a HE $\gamma$-ray flare. This increase in both the number of protons and 
proton-synchrotron photons leads to a strongly enhanced pion (and subsequently, muon) production 
rate. The synchrotron emission from these additional high energy pions and muons produces a 
slightly delayed flare in the VHE $\gamma$-ray bandpass, in tandem with an increased neutrino
flux from pion and muon decay. As in the case of the magnetic-field perturbation, the additional 
secondary electrons/positrons from the pion and muon decay then produce a delayed X-ray synchrotron 
flare. As these secondaries eventually cool down to even lower energies, their synchrotron emission
contributes even to the optical (R-band) flux, leading to a slightly delayed flare in that band. 

The perturbation characterized by an increase of the acceleration efficiency, leads to interesting 
features which are quite distinct from the B-field and injection-luminosity enhancements discussed 
above. With an increase in the stochastic acceleration efficiency, particles diffuse more efficiently
to lower and higher energies. As protons are accelerated to higher energies, the proton synchrotron
emission extends to higher energies, now entering the VHE $\gamma$-ray regime, leading to a prompt
VHE $\gamma$-ray flare. The ultrarelativistic protons interact with the enhanced synchrotron radiation
field, thus increasing the pion and muon production rates. The pions and muons themselves are subject
to the increased acceleration efficiency and are thus pushed to higher energies, leading to a delayed,
secondary VHE $\gamma$-ray flare due to pion and muon synchrotron radiation. All particle distributions 
cool due to synchrotron emission so that the spectral components gradually progress to lower frequencies. 
This leads to delayed flares in the HE $\gamma$-rays as well as X-rays and optical (R-band). 

\begin{figure}[ht]
\begin{center}
\vskip 0.9cm
\includegraphics[height=0.35\textheight]{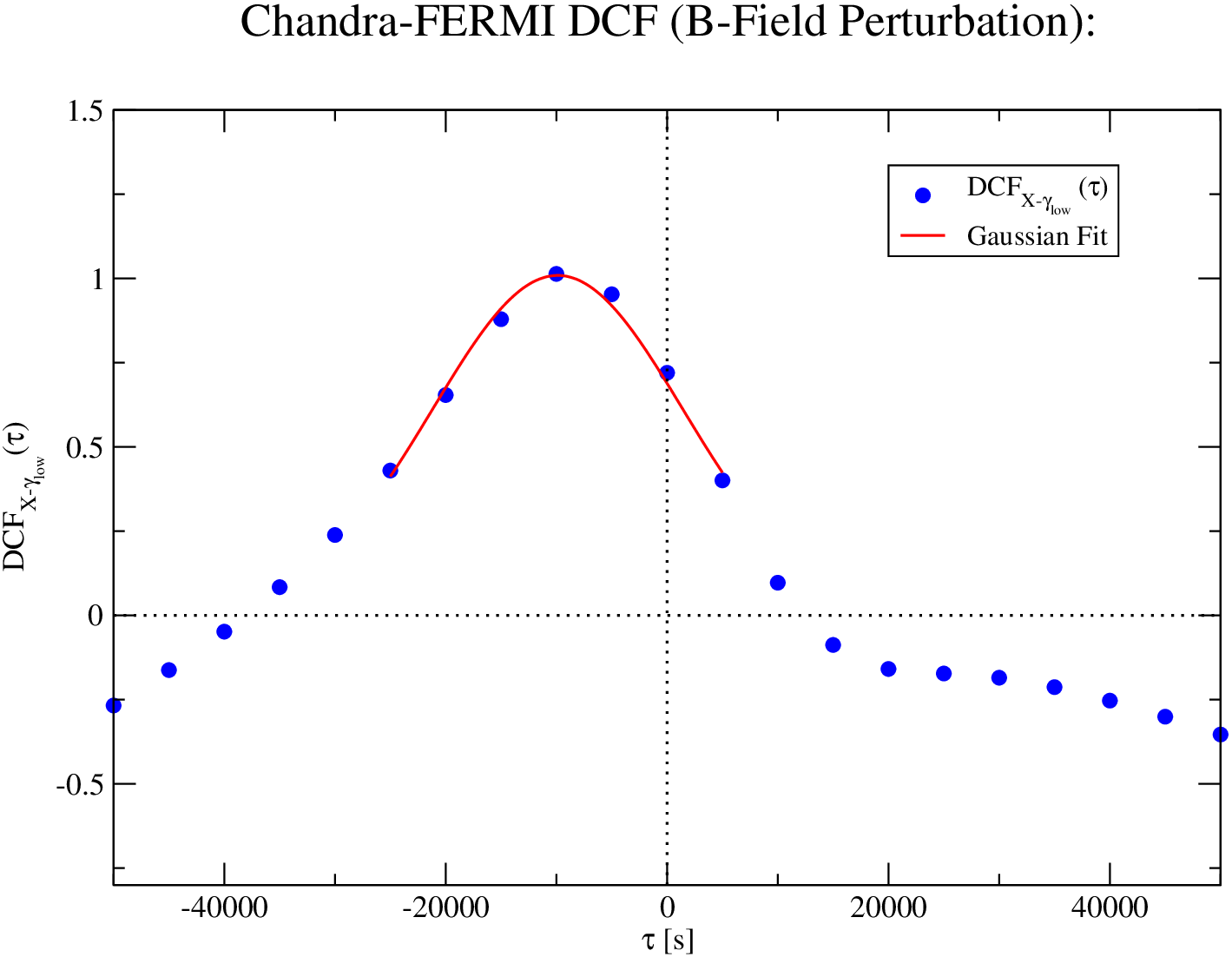}
\caption{The discrete correlation function between the X-ray and HE $\gamma$-ray bandpasses for the 
magnetic field perturbation. A negative time lag indicates that the HE $\gamma$-rays lead the X-rays.}
\label{DCF_XG_lowB}
\end{center}
\end{figure}

The perturbation of the proton spectral index also produces a interesting, distinct features. Due to
the harder proton spectrum, the primary proton synchrotron emission temporarily also makes a larger
contribution to the VHE $\gamma$-ray emission, leading to a brief, pronounced VHE flare. As in the 
case of the other perturbations discussed above, this also leads to increased pion and muon
production rates, leading to delayed X-ray, optical, and neutrino flares. The synchrotron emission
from the cooled, additional highest-energy protons produce a delayed flare in the HE $\gamma$-ray 
bandpass. As the flare progresses, the stronger proton cooling due to pion production results in
a temporarily lower high-energy cut-off of the proton spectrum, because the high-energy cut-off of
the proton injection spectrum remained unchanged, while radiative cooling becomes more efficient. 
As a result, the high-energy end of the proton synchrotron spectrum no longer makes a significant
contribution to the VHE $\gamma$-ray flux, and the pion production rate (and subsequent pion and muon 
synchrotron emission) decreases temporarily. This leads to a dip in the VHE $\gamma$-ray light curve, 
before the perturbation subsides and radiative equilibrium is re-established.

\begin{figure}[ht]
\begin{center}
\vskip 0.9cm
\includegraphics[height=0.35\textheight]{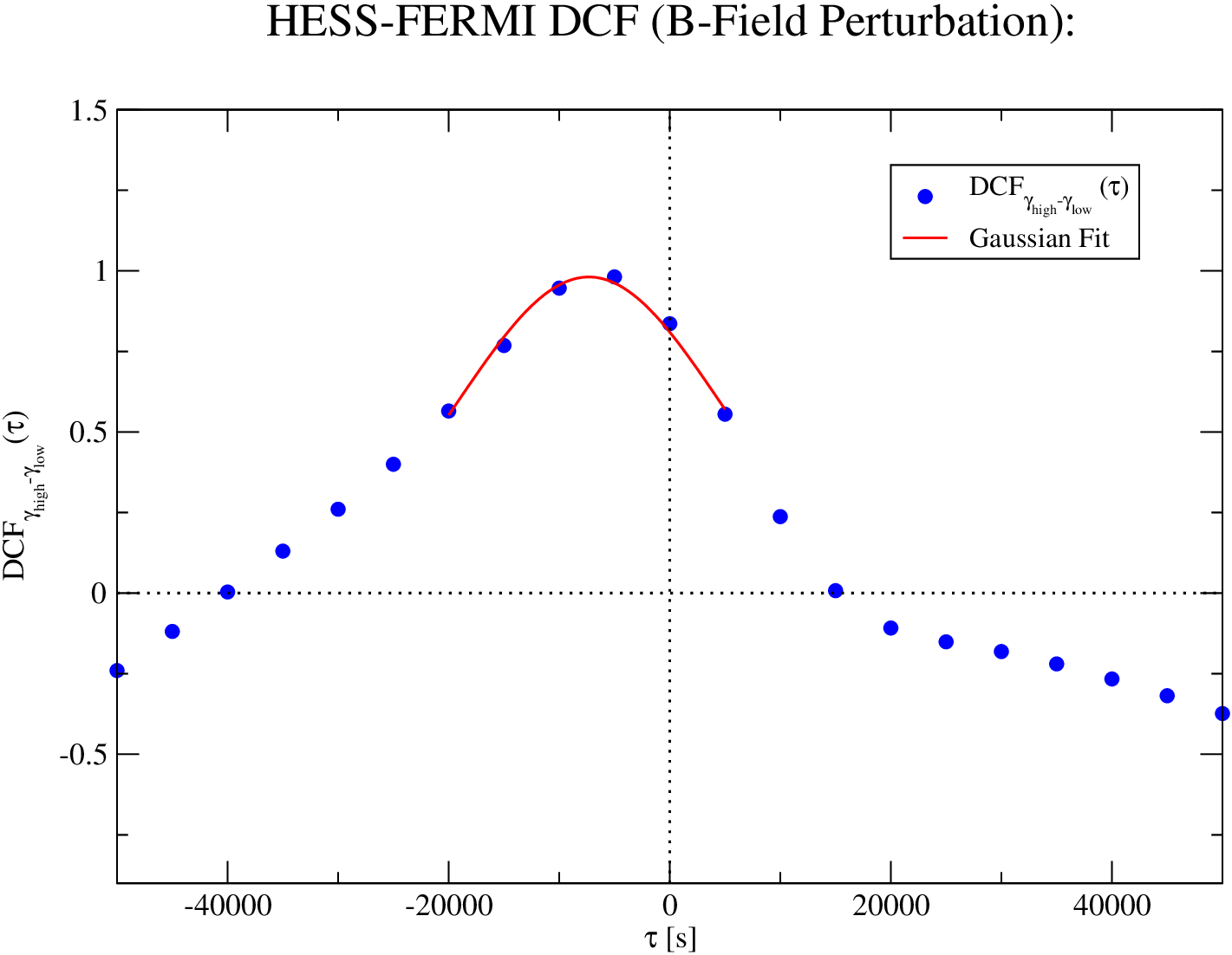}
\caption{The discrete correlation function between the VHE $\gamma$-ray and HE $\gamma$-ray 
bandpasses for the magnetic field perturbation. A negative time lag indicates that the 
HE $\gamma$-rays lead the VHE $\gamma$-rays. }
\label{DCF_G_lowG_highB}
\end{center}
\end{figure}

The distinct features in these lightcurves can be used as a key diagnostic to differentiate 
between one-zone leptonic and hadronic models. In our previous study of the analogous flaring
scenarios in a one-zone leptonic model in \cite{Diltz14}, we predicted that a perturbation 
characterized by an increase in the electron acceleration efficiency would produce a deficit 
in the X-ray emission, while producing marked flares in the R-band, HE and VHE bandpasses. 
In leptonic models to the SEDs of FSRQs (such as 3C279), the X-rays are typically dominated
by the low-energy end of the SSC emission. The drop in the X-ray flux was therefore attributed 
to a shift of the SSC emission to higher energies as a consequence of the increased electron 
acceleration efficiency. In contrast, in hadronic-model fits, the X-rays are dominated by
synchrotron radiation of relativistic protons at intermediate energies ($\gamma_p \sim 10^6$). 
When an acceleration-time-scale perturbation is applied to the hadronic model, all particle 
populations (including protons, pions, and muons) are accelerated to higher energies, without
substantially affecting the particle distributions at intermediate energies. This causes 
increased pion, muon and electron-positron production rates. Following the subsequent
radiative cooling of secondary electrons/positrons, their synchrotron emission leads to 
a delayed X-ray flare. 

Additional marked differences are the delayed VHE $\gamma$-ray plateau found in our simulation
of the acceleration-efficiency perturbation and the dip in the VHE light curve predicted for
the proton spectral-index perturbation, both of which are not expected in leptonic models. 
These marked differences in the multiwavelength light curves may serve as diatnostics to
distinguish between one-zone leptonic and hadronic models.

\begin{figure}[ht]
\begin{center}
\vskip 0.9cm
\includegraphics[height=0.35\textheight]{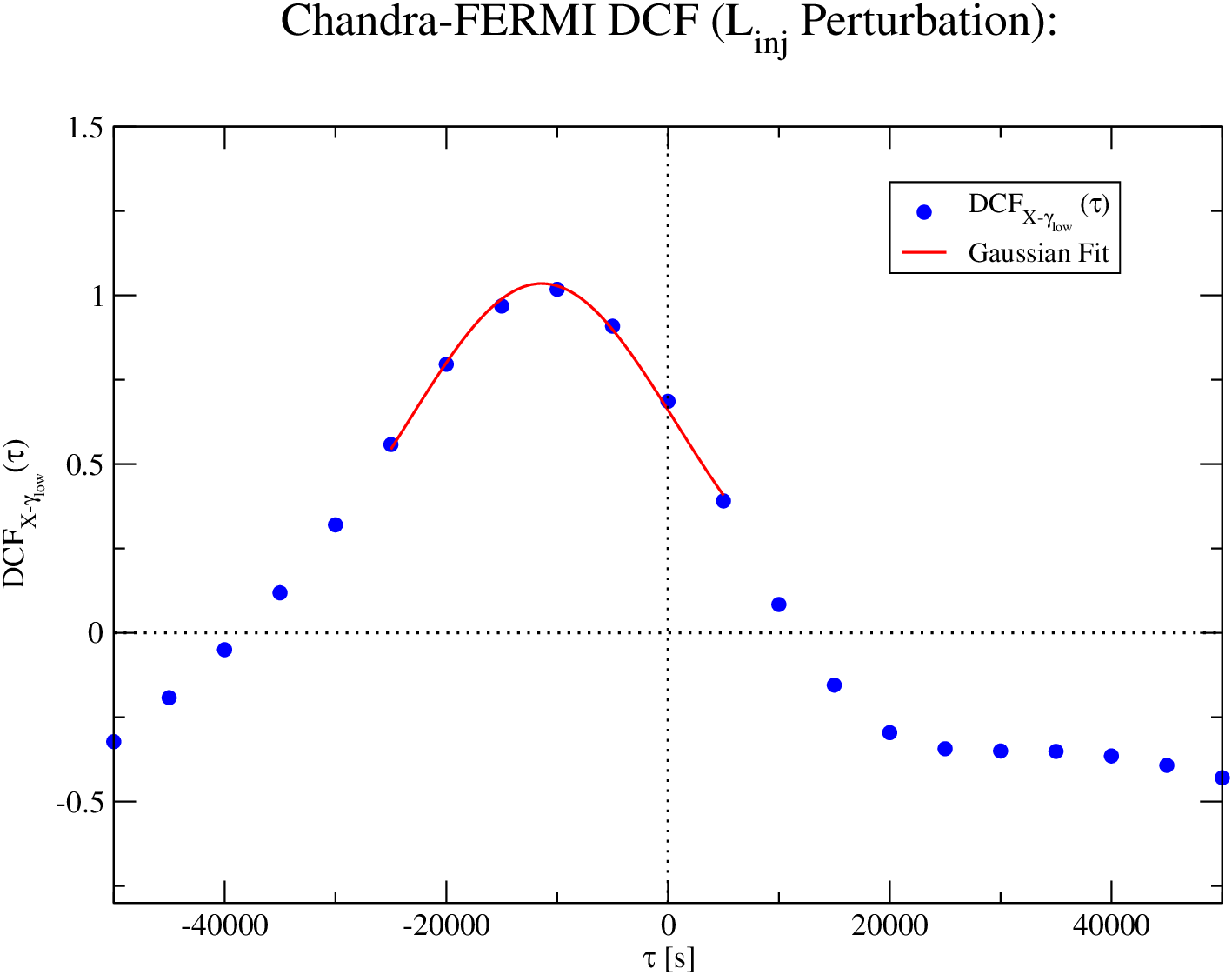}
\caption{The discrete correlation function between the X-ray and HE $\gamma$-ray bandpasses for the 
proton injection perturbation. A negative lag indicates an X-ray lag behind the HE $\gamma$-rays. }
\label{DCF_XG_lowL}
\end{center}
\end{figure}

\section{\label{correlation}Discrete Correlation Analysis:}

In order to further analyze cross-correlations and time lags between the simulated light curves
discussed in the previous section, we calculate the discrete correlation functions \citep[DCF,][]{Edelson88}
between the light curves in the various bandpasses. In order to quantify the preferred values of the strength
of the correlation (the maximum amplitude of the DCF) and inter-band time lag, a Gaussian fit to the DCFs
was performed that minimized the chi-square between the data set and a fitting function of the form

\begin{equation}
    f(\tau) = F_{1} \cdot e^{-(\tau - \tau_{\rm pk})^{2}/2 \sigma^{2}}
\end{equation}

For this purpose, in order to be able to evaluate a $\chi^2$ value, we arbitrarily assumed a relative 
flux error of 1~\% for each simulated light curve point when calculating the DCFs and their errors. 
In this discussion, we focus on the X-ray through $\gamma$-ray portion of the spectrum, and thus, on
the DCFs between X-rays, HE $\gamma$-rays, and VHE $\gamma$-rays. This is largely motivated by significant
differences between leptonic and hadronic models in the X-ray and $\gamma$-ray light curves for the 
acceleration time scale and proton spectral index perturbations. The best fit parameters for the 
various flaring scenarios are listed in Table 3.

\begin{figure}[ht]
\begin{center}
\vskip 0.9cm
\includegraphics[height=0.35\textheight]{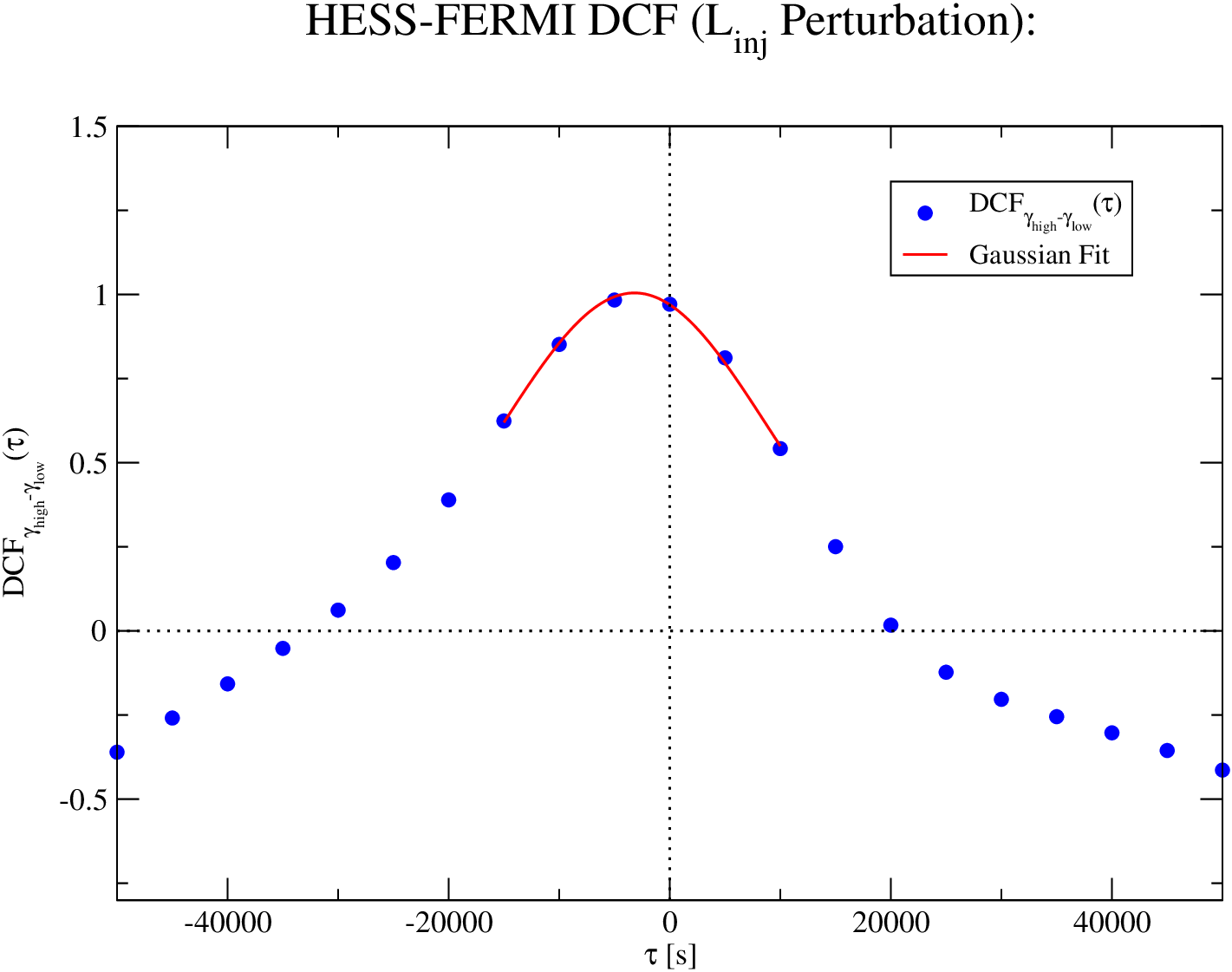}
\caption{The discrete correlation function between the VHE $\gamma$-ray and HE $\gamma$-ray 
bandpasses for the proton injection perturbation. A negative lag indicates a VHE $\gamma$-ray
lag behind the HE $\gamma$-rays. }
\label{DCF_G_lowG_highL}
\end{center}
\end{figure}

The discrete correlation functions show strong correlations between the X-ray, HE, and VHE bandpasses 
for all perturbations considered in this paper. For both the magnetic field and proton injection 
luminosity perturbations, we find that the HE $\gamma$-ray flare is followed by a flare in 
VHE $\gamma$-rays and finally by a flare in the X-ray bandpass. This gives credence to the 
physical scenario discussed in the previous section that an increase in the synchrotron photon 
fields will subsequently generate flares in the VHE $\gamma$-ray bandpass and then a delayed 
X-ray flare.

\begin{figure}[ht]
\begin{center}
\vskip 0.9cm
\includegraphics[height=0.35\textheight]{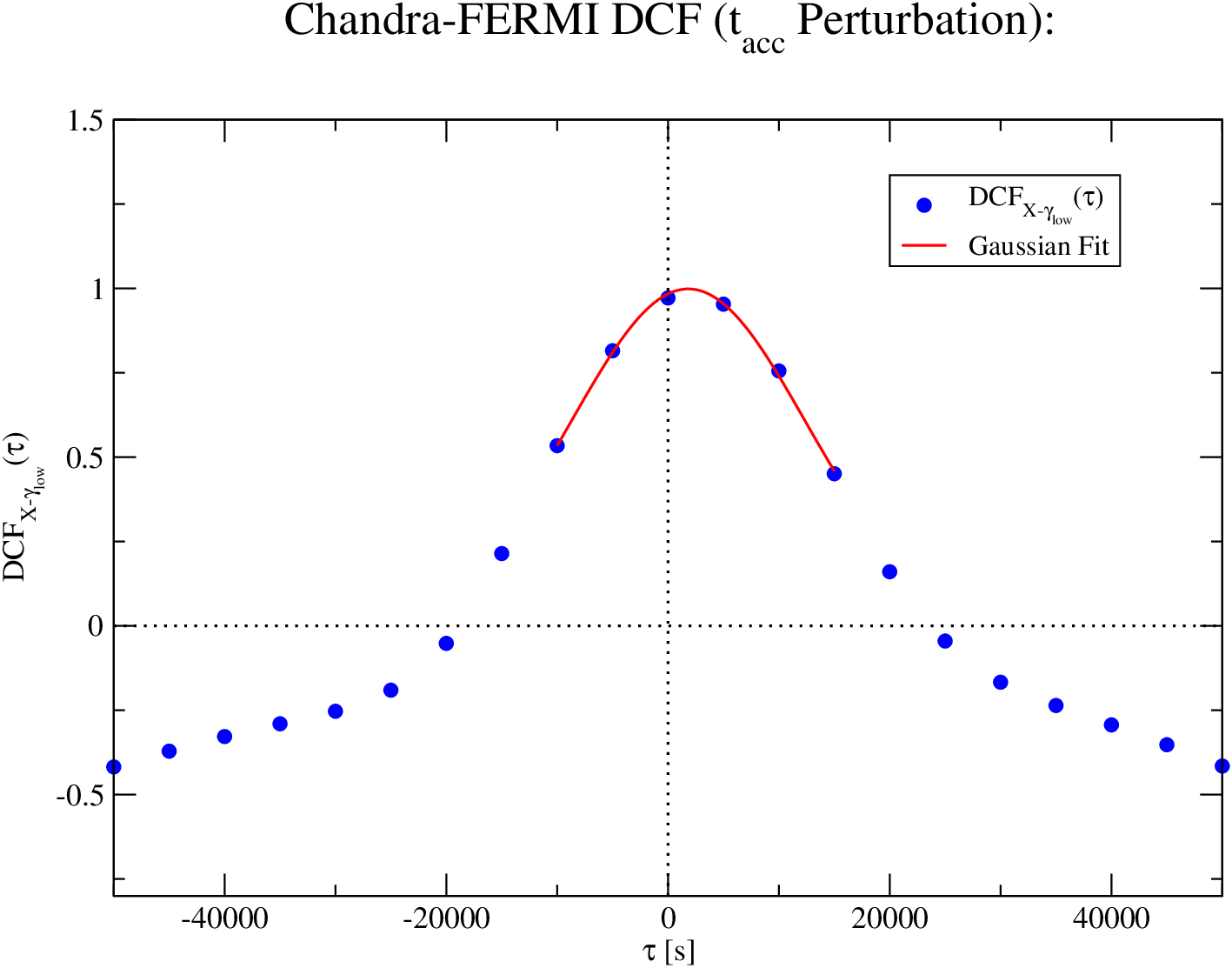}
\caption{The discrete correlation function between the X-ray and HE $\gamma$-ray bandpasses for 
the acceleration time scale perturbation. A negative lag indicates an X-ray lag behind the HE
$\gamma$-rays. }
\label{DCF_XG_lowt}
\end{center}
\end{figure}

For the acceleration timescale and the proton spectral index perturbation, the DCF analysis confirms 
the leading VHE $\gamma$-ray flare, followed by delayed HE $\gamma$-ray and X-ray flares. Time lags
between the X-ray and $\gamma$-ray bands are typically $\sim 1$ -- a few hours. Within error bars, 
the time lags determined from the DCFs agree with those extracted from visual inspection of the 
light curves. Variability on such time scales (and, thus, corresponding inter-band time lags) is 
easily measurable in X-rays and VHE $\gamma$-rays. However, the measurement of HE $\gamma$-ray
variability on time scales of a few hours by {\it Fermi}-LAT may be possible only in extraordinarily 
high flux states.

\begin{figure}[ht]
\begin{center}
\vskip 0.9cm
\includegraphics[height=0.35\textheight]{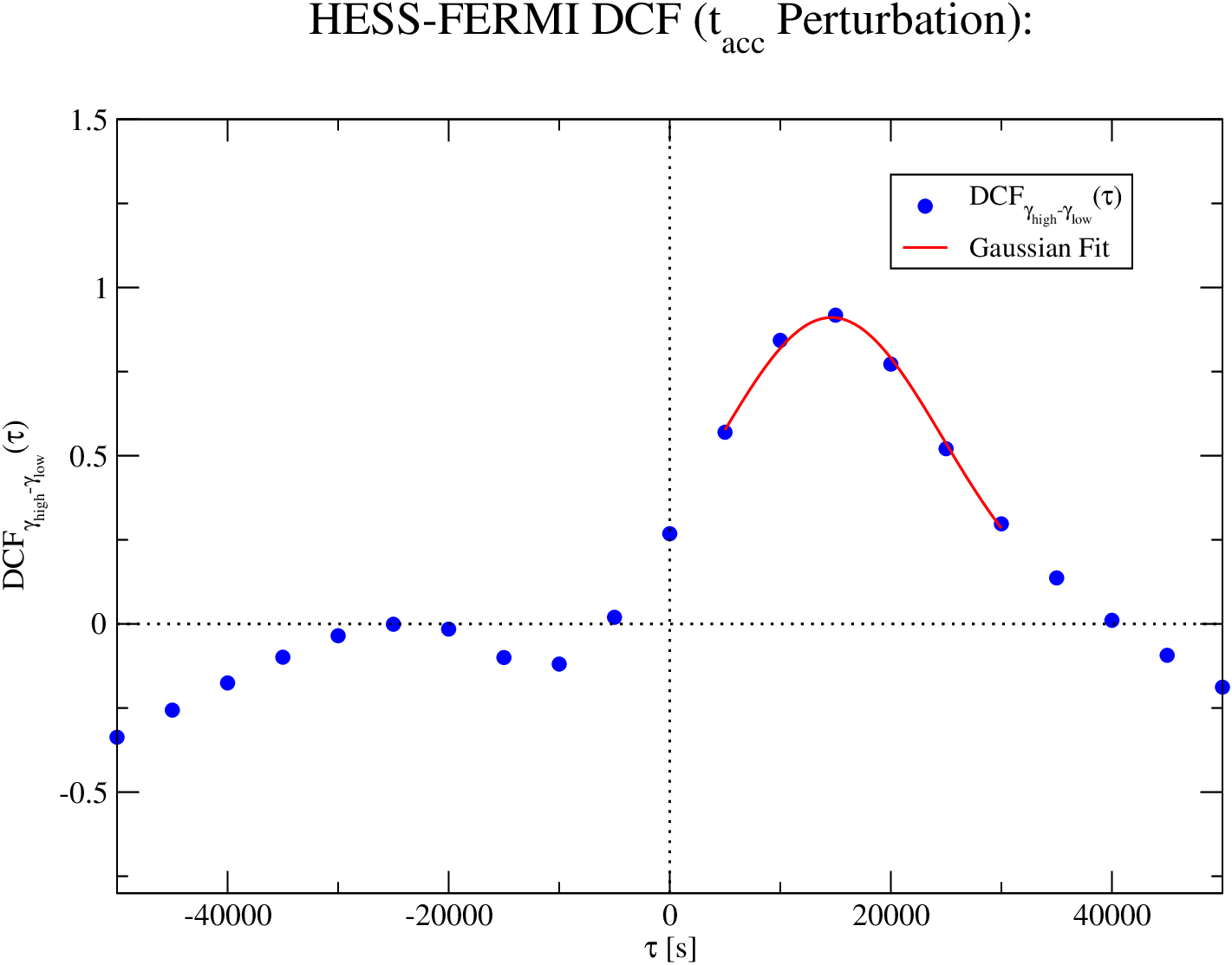}
\caption{The discrete correlation function between the VHE $\gamma$-ray and HE $\gamma$-ray bandpasses 
for the acceleration time scale perturbation. A positive time lag indicates an HE $\gamma$-ray lag
behind the VHE $\gamma$-rays.}
\label{DCF_G_lowG_hight}
\end{center}
\end{figure}

\begin{table}[ht]
\centering
\begin{tabular}{cccccc}
\hline
Bands & Scenario & $F_{1}$ & $ \sigma [s]$ & $\tau_{\rm pk} [s]$ & Fig. \\
\hline
X-HE & $B$  & $1.01$ & $(1.59 \pm 0.13) \times 10^{4}$ & $(-9.88 \pm 1.03) \times 10^{3}$ & \ref{DCF_XG_lowB} \\
HE-VHE & $B$ & $0.98$ & $(1.67 \pm 0.19) \times 10^{4}$ & $(-7.32 \pm 1.29) \times 10^{3}$ & \ref{DCF_G_lowG_highB} \\
X-HE & $L_{\rm p, inj}$  & $1.04$ & $(1.69 \pm 0.15) \times 10^{4}$ & $(-1.14 \pm 0.11) \times 10^{4}$ & \ref{DCF_XG_lowL} \\
HE-VHE & $L_{\rm p, inj}$ & $1.01$ & $(1.69 \pm 0.21) \times 10^{4}$ & $(-3.19 \pm 1.34) \times 10^{3}$ & \ref{DCF_G_lowG_highL} \\
X-HE & $t_{\rm acc}$ & $0.99$ & $(1.49 \pm 0.17) \times 10^{4}$ & $(1.83 \pm 1.19) \times 10^{3}$ & \ref{DCF_XG_lowt} \\
HE-VHE & $t_{\rm acc}$ & $0.91$ & $(1.42 \pm 0.18) \times 10^{4}$ & $(1.46 \pm 0.14) \times 10^{4}$ & \ref{DCF_G_lowG_hight} \\
X-HE & $q_{p}$  & $0.99$ & $(3.39 \pm 0.63) \times 10^{4}$ & $(-4.08 \pm 2.31) \times 10^{3}$ & \ref{DCF_XG_lowq} \\
HE-VHE & $q_{p}$ & $0.87$ & $(3.43 \pm 1.05) \times 10^{4}$ & $(2.64 \pm 0.53) \times 10^{3}$ & \ref{DCF_G_lowG_highq} \\
\hline
\end{tabular} \\ 
\caption{Best-fit DCF correlation strengths and time lags from the Gaussian fits to the discrete correlation 
functions.}
\label{Gaussian_parameters}
\end{table}

\begin{figure}[ht]
\begin{center}
\vskip 0.9cm
\includegraphics[height=0.35\textheight]{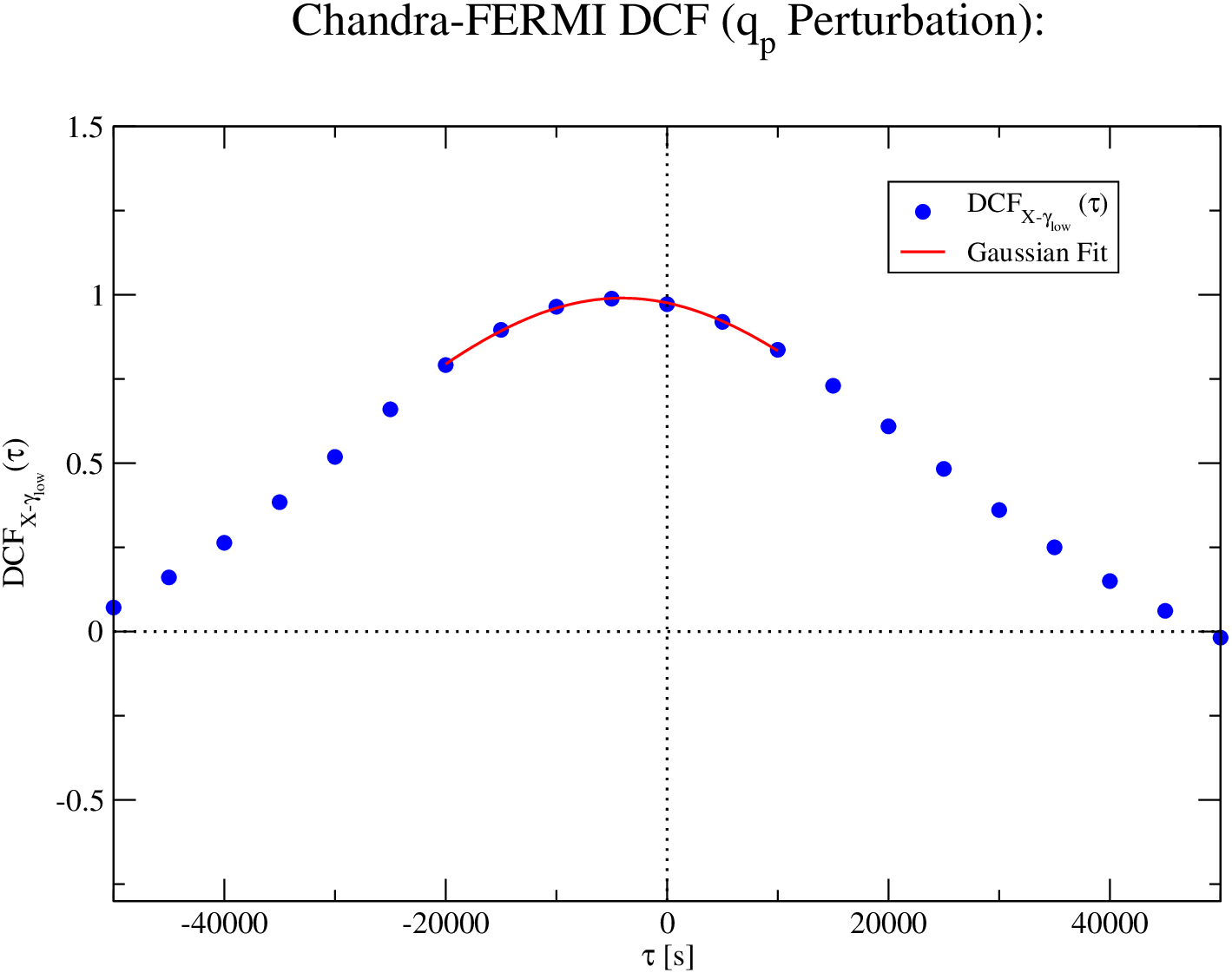}
\caption{The discrete correlation function between the X-ray and HE $\gamma$-ray bandpasses for the 
proton spectral index perturbation. A negative lag indicates an X-ray lag behind the HE $\gamma$-rays. }
\label{DCF_XG_lowq}
\end{center}
\end{figure}

\begin{figure}[ht]
\begin{center}
\vskip 0.9cm
\includegraphics[height=0.35\textheight]{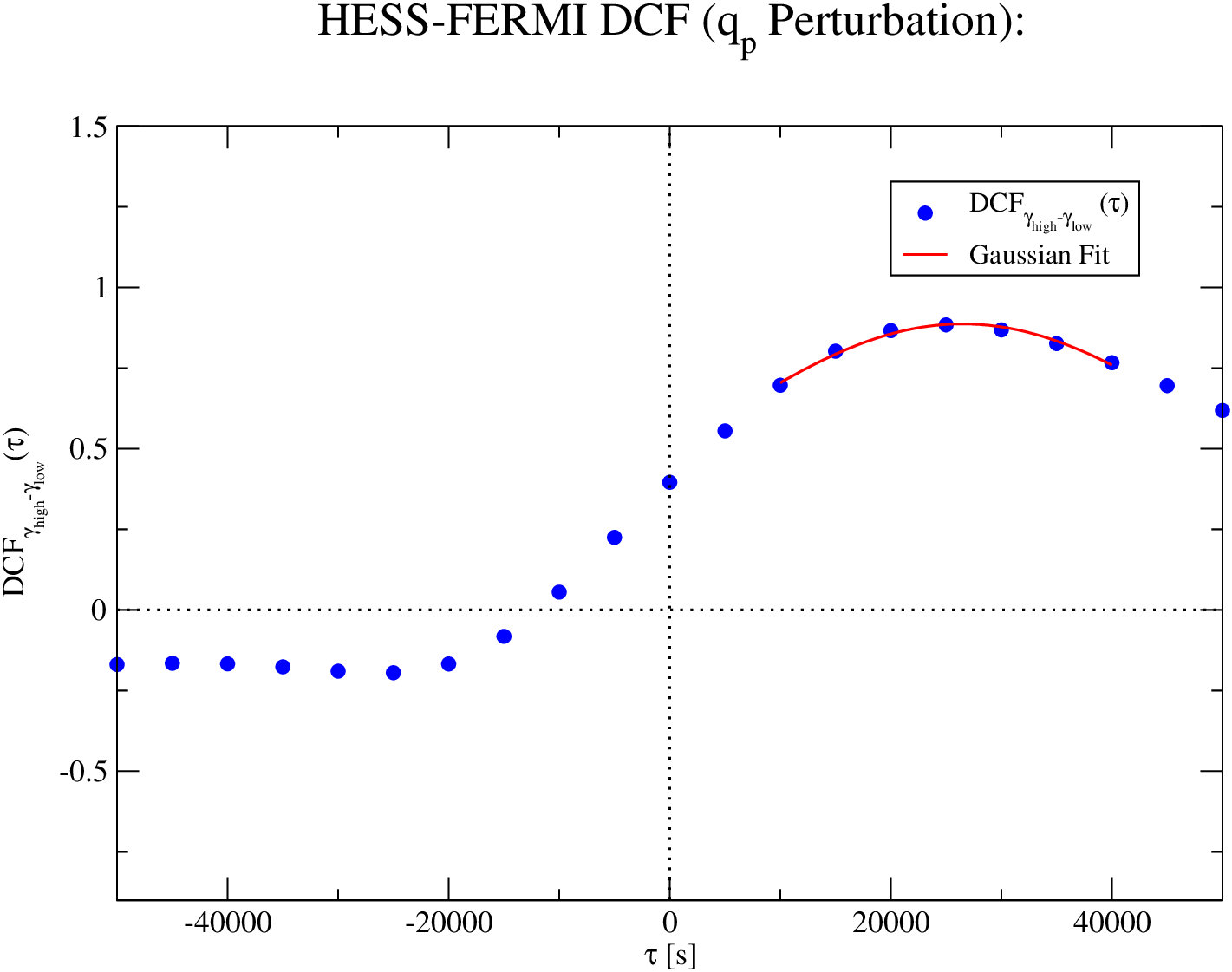}
\caption{The discrete correlation function between the VHE $\gamma$-ray and HE $\gamma$-ray bandpasses 
for the proton spectral index perturbation. A positive lag indicates a HE $\gamma$-ray lag behind the 
VHE $\gamma$-rays. }
\label{DCF_G_lowG_highq}
\end{center}
\end{figure}

\section{\label{conclusions}Summary and Conclusions:}

In this paper, we have described a new time-dependent lepto-hadronic model for the broadband emission
of relativistic jet sources (especially blazars) that incorporates the synchrotron radiation from 
secondary pions and muons, generated in photo-hadronic interactions. We use an equilibrium solution 
of our model to produce a rough fit-by-eye to the high-state SED of the FSRQ 3C 279. The broadband 
emission from infra-red to VHE $\gamma$-rays can be explained by a combination of synchrotron emission 
from electrons, protons, pions, and muons. The parameters used for the fit are chosen so that 
the radiative contributions from the pion and muon synchrotron radiation are non-negligible, 
and their contribution is essential for an adequate fit of the unusually hard VHE $\gamma$-ray 
spectrum measured by MAGIC.  Our fits for 3C 279 can be achieved with the jet being close to
equipartition between the power carried in magnetic fields (Poynting flux) and the kinetic energy
in protons and electrons ($\epsilon_B \equiv L_B / L_{\rm kin} = 0.29$). This contrasts most other 
hadronic models in which the particle kinetic luminosity is a few orders of magnitude larger than 
the magnetic luminosity. 
However, our model requires similarly large jet powers as other published hadronic blazar model
fits \citep[e.g.,][and references therein]{Boettcher13}, greatly in excess of the Eddington luminosity 
of the central black hole in 3C279 ($L_{\rm Edd} \sim 8 \times 10^{46}$~erg~s$^{-1}$). Nevertheless, as 
the jet power is strongly beamed perpendicular to the accretion flow, it does not provide a radiation
pressure that would be able to shut off the accretion flow. Therefore, the Eddington limit argument
may not apply in such a case. Nevertheless, the extreme jet production efficiency required for 
hadronic blazar jet models in general, may constitute a problem for this kind of models: The
total jet power of $L_j \sim 1.3 \times 10^{49}$~erg~s$^{-1}$ exceeds the radiative luminosity
of the accretion disk ($L_d \sim 2 \times 10^{45}$~erg~s$^{-1}$) by almost 4 orders of magnitude.
No jet production mechanism is currently known that would be able to produce steady jets with this
efficiency; however, the current understanding of the production of relativistic jets from supermassive 
black holes is still very limited. This issue is further discussed in \cite{zb15}.

We have then simulated light curves by applying perturbations to a various input parameters in 
our code. The perturbations of the magnetic field and proton injection luminosity produced strong 
correlations between all bandpasses with $3 - 4$ hour time lags between the HE $\gamma$-ray and X-ray 
bandpass and $1 - 2$ hour time lags between the VHE and HE $\gamma$-ray bandpasses. Also a temporary 
increase in the stochastic acceleration efficiency leads to correlated flares in the $\gamma$-ray and
X-ray bandpasses. This is in contrast to the the effects of such a perturbation on a time-dependent
leptonic model \citep{Diltz14}, in which a drop in X-ray emission was predicted. The predicted 
variability features are well within reach of observational capabilities of currently operating
X-ray and VHE $\gamma$-ray observatories, but require extraordinarily high flux states to be
measurable by {\it Fermi}-LAT. 
Our baseline (quiescent-state) model fit simulations predict integrated neutrino number 
fluxes at Earth, over the IceCube energy range for all three neutrino species, of 
$\approx 10^{-16}$~cm$^{-2}$~s$^{-1}$. Given IceCube's effective area of 
$A_{eff} ( > 100 {\rm TeV}) \approx 10^{8}$~cm$^{2}$, this predicts neutrino detection
rates of $\sim 0.3$~yr$^{-1}$, thus requiring $\gtrsim 10$~years for a significant detection of neutrinos
from 3C279 in quiescence. Even during flaring states, as studied in this paper, the neutrino flux is
expected to increase by factors of a few -- a few tens, to expected detection rates of $\sim 10^{-7}$~s$^{-1}$, 
rendering it unlikely that IceCube would be able to detect a neutrino signal correlated with $\gamma$-ray
flares from 3C279.   

The most interesting features in our simulated lightcurve were plateaus and dips in the VHE $\gamma$-ray 
bandpass as a result of perturbations of the acceleration time scale or the proton injection spectral 
index. These plateaus are primarily caused by delayed synchrotron radiation from the secondary 
pions and muons. Such VHE light curve plateaus / dips are not predicted in one zone leptonic models 
and could be a tell tale signature of hadronic emission from blazar jets in parameter regimes in
which muon and pion synchrotron emission is non-negligible.

\section*{Acknowledgments}

This work was funded by NASA through Astrophysics Data Analysis Program grant NNX12AE43G.
The work of M.B. is supported through the South African Research Chair Initiative
of the National Research Foundation and the Department of Science and Technology
of South Africa, under SARChI Chair grant No. 64789.


\begin{thebibliography}{}

\bibitem[Abdo et al.(2010)]{Abdo10}
Abdo, A. A., et al., 2010, 716, 30

\bibitem[Aharonian et al.(2000)]{Aharonian00}
Aharonian, F. A., 2000, New Astron., 5, 377

\bibitem[Albert et al.(2008)]{Albert08}
Albert, J., et al., 2008, Science, 320, 1752

\bibitem[Barr et al.(1988)]{Barr88}
Barr, S., Gaisser, T. K., et al., 1988, Phys. Letters 214, 147.

\bibitem[Ballo et al.(2002)]{Ballo02}
Ballo, L., Maraschi, L., et al., 2002, ApJ, 567, 50B

\bibitem[Bentz et al.(2013)]{Bentz13}
Bentz, M. C., Denney, K. D., et al., 2013, ApJ, 767, 149

\bibitem[Blandford \& Levinson(1995)]{Blandford95}
Blandford, R. D. \& Levinson, A., 1995, ApJ, 441, 79

\bibitem[Blazejowski et al.(2000)]{Blazejowski00}
Blazejowski, M., et al., 2000, ApJ, 545, 107

\bibitem[B\"ottcher \& Schlickeiser(1997)]{2Boettcher97}
B\"ottcher, M., Schlickeiser, R., 1997, A\&A, 325, 866 

\bibitem[B\"ottcher \& Chiang(2002)]{Boettcher02}
B\"ottcher, M., Chiang, J., 2002, ApJ, 581, 127

\bibitem[B\"ottcher et al.(2007)]{Boettcher07}
B\"ottcher, M., et al., 2007, ApJ, 670, 968

\bibitem[B\"ottcher et al.(2009)]{Boettcher09}
B\"ottcher, M., Reimer, A., \& Marscher, A. P., 2009, ApJ, 703, 1168

\bibitem[B\"ottcher et al.(2012)]{Boettcher12}
B\"{o}ttcher, M., Harris, D. E., Krawczynski, H., 2012, 
\textit{"Relativistic Jets from Active Galactic Nuclei"} WILEY-VCH Verlag GmBH \& Co. 

\bibitem[B\"ottcher et al.(2013)]{Boettcher13}
B\"ottcher, M., Reimer, A., Sweeney, K., \& Prakash, A., 2013, ApJ, 768, 54

\bibitem[Dermer et al.(1992)]{Dermer92}
Dermer, C. D., Schlickeiser, R., \& Mastichiadis, A., 1992, A\&A, 256, L27

\bibitem[Dermer \& Schlickeiser(1993)]{Dermer93}
Dermer, C. D., Schlickeiser, R., 1993, ApJ, 416, 458

\bibitem[Dermer \& Menon(2009)]{Dermer09}
Dermer, C. D., Menon, G., 2009, \textit{"High Energy Radiation from Black Holes"} Princeton University Press

\bibitem[Diltz \& B\"ottcher(2014)]{Diltz14}
Diltz, C. \& B\"ottcher, M., 2014, JHEAP, 63D

\bibitem[Edelson et al.(1988)]{Edelson88}
Edelson, R. A., Krolik, J. H., 1988, ApJ, 333, 646

\bibitem[Hartman et al.(1996)]{Hartman96}
Hartman, R. C., et al., 1996, ApJ, 461, 698

\bibitem[Hartman et al.(2001)]{Hartman01}
Hartman, R. C. et al., 2001, ApJ, 553, 683

\bibitem[Hovatta et al.(2009)]{Hovatta09}
Hovatta, T., Valtaoja, E., Tornikoski, M., \& L\"ahteenm\"ki, A., 2009, A\&A, 498, 723

\bibitem[H\"ummer et al. (2010)]{Huemmer10}
H\"ummer, S., R\"uger, M., Spainer, F., Winter, W., 2010, ApJ 721 630H 

\bibitem[Jones(1968)]{Jones68}
Jones, F. C., 1968, Phys. Rev., 167, 1159

\bibitem[Kelner \& Aharonian(2008)]{Kelner08}
Kelner, S. R., Aharonian, F. A., 2008, Phys. Rev. D78, 034013

\bibitem[Mannheim \& Biermann(1992)]{MB92}
Mannheim. K., \& Biermann, P. L., 1992, A\&A, 253, L21

\bibitem[Mannheim(1993)]{Mannheim93}
Mannheim, K., 1993, A\&A, 269, 69

\bibitem[Maraschi et al.(1994)]{Maraschi94}
Maraschi, L., et al., 1994, ApJ, 435, L91

\bibitem[Mastichiadis \& Kirk(1995)]{Mastichiadis95}
Mastichiadis, A., \& Kirk. J. G., 1995, A\&A, 295, 613

\bibitem[Mastichiadis \& Kirk(2005)]{Mastichiadis05}
Mastichiadis, A., \& Kirk. J. G., 2005, A\&A, 433, 765

\bibitem[M\"ucke \& Protheroe(2001)]{Muecke01}M\"ucke, A., \& Protheroe, R. J., 2001, APh, 15, 121

\bibitem[M\"ucke et al.(2003)]{Muecke03}M\"ucke, A., Protheroe, R. J., Engel, R., Rachen, J. P.,
\& Stanev, T., 2003, APh, 18, 593

\bibitem[Schlickeiser(1984a)]{1Schlickeiser84}
Schlickeiser, R., 1984a, A\&A, 136, 227

\bibitem[Schlickeiser(2002)]{Schlickeiser02}
Schlickeiser, R., 2002, Astronomy \& Astrophysics Library

\bibitem[Sikora et al.(1994)]{Sikora94}
Sikora, M., Begelman, M., \& Rees, M., 1994 ApJ, 421, 153

\bibitem[Urry(1998)]{Urry98}
Urry, C. M., 1998, Advances in Space Research, 21 ,89

\bibitem[Weidinger \& Spanier(2015)]{Weidinger14}
Weidinger, M., Spanier, F., 2015, A\&A, 573, A7 

\bibitem[Woo et al.(2002)]{Woo02}
Woo, J. H., Urry, C. M., 2002, ApJ, 579, 530W

\bibitem[Xie et al.(2008)]{Xie08}
Xie, Z. H., Hao, J. M., Du, L. M., Zhang, X., \& Jia, Z. L., 2008, PSAP, 120, 477

\bibitem[Zdziarski \& B\"ottcher(2015)]{zb15}
Zdziarski, A. A., \& B\"ottcher, M., 2015, MNRAS, submitted (arXiv:1501.06124)

\bibitem[Zhang \& B\"ottcher(2013)]{Zhang13}
Zhang, H., \& B\"ottcher, M., 2013, ApJ, 774, 18

\end{thebibliography}
\end{document}